\documentclass[reprint,amsmath,amssymb,aps,superscriptaddress,twocolumn,nofootinbib,10pt]{revtex4-1}
\usepackage[colorlinks=true,allcolors=blue]{hyperref}
\usepackage{graphicx,color}
\usepackage{dcolumn}
\usepackage{subfigure}
\usepackage{bm}
\usepackage{amsmath,amsfonts,amssymb}
\usepackage{acronym}
\usepackage{enumitem}
\usepackage{array}
\usepackage{multirow}
\usepackage{xcolor}
\usepackage{CJK}
\usepackage[normalem]{ulem}

\allowdisplaybreaks
\newcommand{\be}{\begin{equation}}
\newcommand{\ee}{\end{equation}}
\newcommand{\bea}{\begin{eqnarray}}
\newcommand{\eea}{\end{eqnarray}}

\newcommand{\NCUa}{Department of Physics, Nanchang University, Nanchang, 330031, China}
\newcommand{\NCUb}{Center for Relativistic Astrophysics and High Energy Physics, Nanchang University, Nanchang, 330031, China}

\newacro{EMRI}{extreme mass-ratio inspirals}
\newacro{MBH}{massive black hole}
\newacro{BH}{black hole}
\newacro{GR}{general relativity}
\newacro{HKBH}{hairy Kerr black hole}
\newacro{KNBH}{Kerr-Newmann black hole}
\newacro{KBH}{Kerr black hole}
\newacro{NHT}{no-hair theorem}
\newacro{DWD}{double white dwarf}
\newacro{GW}{gravitational wave}
\newacro{AK}{analytic kludge}
\newacro{NK}{numerical kludge}
\newacro{AAK}{augmented analytic kludge}
\newacro{CO}{compact object}
\newacro{PE}{parameter estimation}
\newacro{SNR}{signal-to-noise ratio}
\newacro{PN}{post newtonion}
\newacro{FIM}{Fisher information matrix}
\newacro{LSO}{last stable orbit}
\newacro{ISCO}{innermost stable circular orbit}
\newacro{BBH}{Binary Black Hole}
\newacro{BNS}{Binary Neutron Star}
\newacro{NS}{Neutron Star}
\newacro{KN}{Kerr-Newmann}
\newcommand{\beq}{\begin{equation}}
\newcommand{\eeq}{\end{equation}}
\newcommand{\beqa}{\begin{eqnarray}}
\newcommand{\eeqa}{\end{eqnarray}}

\def\lsim{\mathrel{\rlap{\lower4pt\hbox{\hskip0.5pt$\sim$}}
    \raise1pt\hbox{$<$}}}         %less than or approx. symbol
\def\gsim{\mathrel{\rlap{\lower4pt\hbox{\hskip0.5pt$\sim$}}
    \raise1pt\hbox{$>$}}}         %greater than or approx. symbol
\begin{document}
\begin{CJK*}{UTF8}{gbsn}
\title{{\Large {\bf Preliminary forecasting constraint on scalar charge with LISA in non-vacuum environments}}}

\author{Tieguang Zi}
\email{zitieguang@ncu.edu.cn}
\affiliation{\NCUa}
\affiliation{\NCUb}

\author{Chang-Qing Ye}
\email{yechangqing@suse.edu.cn, corresponding author}
\affiliation{School of Physics and Electronic Engineering, Sichuan University of Science $\&$ Engineering, Zigong 643000, People's Republic of China}

\begin{abstract}
We compute the gravitational wave signal from eccentric extreme-mass-ratio inspirals (EMRIs) embedded within beyond-vacuum environments, where the secondary object carries a scalar charge and evolves in the presence of both an accretion disk and a dark matter halo. The waveform modification is derived by incorporating the scalar charge correcting the fluxes and orbital trajectories of the secondary. Our results indicate that, under suitable parameter configurations, the influence of the scalar charge on EMRIs waveform in such environments can be distinguished from that in vacuum spacetime. For the EMRIs signal modified by the astrophysical environments, the future space-borne detector can determine the relative error of scalar charge constrained by LISA at the level of $\sim0.1$, providing a preliminary prediction of detecting scalar charge in the beyond-vacuum spacetime.
\end{abstract}
\maketitle
\end{CJK*}

\section{Introduction}
To date, over one hundred gravitational-wave (GW) events have been detected by the LIGO-Virgo-KAGRA (LVK) Collaboration, providing a wealth of information about the fundamental physics of compact objects~\cite{LIGOScientific:2016aoc,KAGRA:2021vkt,LIGOScientific:2020ibl}. These observations have enabled stringent tests of gravity in the strong-field regime~\cite{LIGOScientific:2016lio,LIGOScientific:2021sio,LIGOScientific:2018dkp,LIGOScientific:2020tif}. So far, however, no statistically significant deviation from general relativity (GR) has been identified in the LVK data. This null result may be partly attributed to the sensitivity limitations of current ground-based detectors, whose low-frequency performance is constrained by seismic noise.

Forthcoming space-based GW observatories, such as the Laser Interferometer Space Antenna (LISA), will drastically expand the accessible source population, particularly for massive compact objects. Among LISA's primary targets are extreme-mass-ratio inspirals (EMRIs), in which a stellar-mass compact object inspirals into a massive black hole (MBH), with a characteristic mass-ratio $\eta \in [10^{-7},10^{-4}]$~\cite{Berry:2019wgg,LISA:2022yao}. Typical EMRIs consist of a secondary with mass $\mu \sim 1$-$10^2,{\rm M}_\odot$ orbiting a MBH of mass $M \sim 10^4$-$10^6,{\rm M}_\odot$. Over the course of the inspiral, the secondary can undergo $\sim 10^4$-$10^5$ orbital cycles before plunge. The accumulation of such a large number of cycles renders EMRIs exquisitely sensitive probes of the spacetime geometry in the immediate vicinity of MBHs~\cite{Barack:2018yly,Barack:2006pq,Babak:2017tow,Fan:2020zhy,Zi:2021pdp,LISA:2022kgy}.

Beyond their role in mapping black-hole spacetimes, EMRIs are also promising laboratories for studying astrophysical environments~\cite{Barausse:2014tra,Hannuksela:2018izj,Cardoso:2019upw,Toubiana:2020drf,DeLuca:2022xlz,Cole:2022yzw,Dyson:2025dlj,Becker:2024ibd,Pan:2021ksp,Zwick:2025wkt,Kejriwal:2023djc}. In particular, they can be used to probe dark-matter (DM) distributions around MBHs~\cite{Eda:2013gg,Eda:2014kra,Macedo:2013qea,Yue:2017iwc,Giudice:2016zpa,Kavanagh:2020cfn,Chakraborty:2024gcr,Becker:2021ivq,Li:2021pxf,Becker:2022wlo,Berezhiani:2023vlo,Bhalla:2024lta,Karydas:2024fcn,Zhao:2024bpp,Boudon:2023mdh,Dai:2023cft} and accretion disks~\cite{Barausse:2007dy,Kocsis:2011dr,Yunes:2011ws,McKernan:2014oxa,Derdzinski:2018qzv}. Such non-vacuum effects can modify both the generation and the detection of GWs by imprinting characteristic signatures on the waveform. Developing accurate waveform models that incorporate environmental corrections is therefore essential to fully exploit EMRIs as probes of both massive compact objects and their surroundings~\cite{Cardoso:2022whc,Duque:2023seg,Berti:2015itd,Cardoso:2019rvt,Spieksma:2024voy}.

Substantial progress has been made in modeling EMRIs orbital dynamics and waveform generation in vacuum GR over the past several decades, driven by advances in self-force theory and related formalisms~\cite{Barack:2018yvs,Hughes:2021exa}, as well as by the development of efficient waveform-generation frameworks and numerical tools~\cite{Chua:2020stf,Katz:2021yft,Chapman-Bird:2025xtd}. Nevertheless, existing waveform families remain incomplete, and further refinement is required to meet the accuracy requirements of LISA data analysis. Extending these models to include environmental effects is even more challenging. Among the difficulties are the absence of fully relativistic rotating black-hole solutions consistently coupled to realistic DM distributions, and the complicated interplay between the spacetime background and matter fields. As a result, many existing studies of environmental signatures in EMRIs rely on post-Newtonian approximations and often neglect the dissipative impact of radiation reaction in non-vacuum spacetimes
~\cite{Kocsis:2011dr,Kavanagh:2020cfn,Coogan:2021uqv,Cole:2022yzw,Tomaselli:2023ysb,Berezhiani:2023vlo}.

Astrophysical environments are known to play a crucial role in the dynamics of binary systems and have attracted growing attention in the context of space-based GW astronomy. Because EMRIs can sever as precision probes of the spacetime geometry near MBHs, they are expected to be particularly sensitive to environmental effects, including accretion disks~\cite{Kocsis:2011dr,Zwick:2021dlg,Caputo:2020irr,Speri:2022upm,Pan:2021oob,Li:2025zgo,Lyu:2024gnk} and DM structures~\cite{Speeney:2022ryg,Cardoso:2022whc,Cardoso:2021wlq,Cannizzaro:2023jle,Aurrekoetxea:2023jwk,Rahman:2023sof,Duque:2023seg,Tomaselli:2024ojz}. These environments can leave distinctive imprints on the energy and angular-momentum fluxes, thereby altering the orbital evolution and, consequently, the observable waveform~\cite{Zwick:2022dih,Cole:2022yzw,Kejriwal:2023djc,Cardoso:2022whc,Duque:2024mfw,Yuan:2024duo}. In parallel, EMRIs provide an exceptional arena for testing fundamental physics. The increasing number of works has explored how LISA observations of EMRIs could be used to place tight constraints on new degrees of freedom, such as scalar fields~\cite{Maselli:2020zgv,Maselli:2021men,Barsanti:2022ana,Barsanti:2022vvl,DellaRocca:2024pnm,Barsanti:2024kul,Speri:2024qak,Jiang:2021htl,Guo:2022euk,Zhang:2022rfr}, vector fields~\cite{Zhang:2022hbt,Zhang:2024ogc,Zi:2024lmt}, and tensor fields~\cite{Cardoso:2018zhm}. These additional fields are expected to imprint characteristic signatures on EMRI GWs, thereby offering a unique opportunity to constrain their properties, such as masses and charges. Thus, the continued study of EMRIs not only advances GW astrophysics but also sheds light on the fundamental structure of spacetime and possible extensions of GR.

In this work, we investigate the prospects for constraining a scalar charge within a class of modified gravity theories using EMRIs evolving in a non-vacuum environment. Concretely, we consider a scalar-tensor framework in which the spacetime of the central MBH remains described by the no-hair theorem, while the secondary can acquire an effective scalar charge. This setup has been extensively discussed in Refs.~\cite{Maselli:2020zgv,Maselli:2021men,Barsanti:2022ana}. The non-vacuum background is modeled as a Schwarzschild MBH surrounded by a Newtonian, stationary, thin accretion disk and an encompassing dark-matter minispike. This prescription is admittedly idealized; it provides a simplified beyond-vacuum spacetime intended to yield preliminary estimates of the capability of EMRIs to bound scalar charges in complex astrophysical environments. A fully consistent treatment would require constructing a relativistic framework that simultaneously models the accretion disk, DM drag, and scalar radiation in a generic beyond-GR theory. Developing such a framework is an important goal for future work.

The remainder of this paper is organized as follows. In Sec.~\ref{env:mehod} we introduce the environmental effects (accretion disk and DM friction), describe the scalar-charged secondary in EMRIs, and outline our scheme for evolving the system. Section~\ref{wave:data:analysis} presents the quadrupole formulas used for waveform generation and the data-analysis methodology. In Sec.~\ref{result} we show how environmental effects modify EMRI waveforms, quantify the interplay between scalar emission and environmental corrections, and derive projected constraints on the scalar charge using the Fisher information matrix (FIM). Finally, we summarize our findings and discuss future directions in Sec.~\ref{discussion}.

\section{Theoretical models}\label{env:mehod}
In this section, we model the motion of the secondary in the environment surrounding MBH, focusing on two representative scenarios: an accretion-disk background and a DM halo. For each case, we compute the corresponding modifications to the EMRI energy and angular-momentum fluxes and quantify how these altered fluxes feed back into the long-duration evolution of the orbital parameters.
\subsection{Accretion-disk model}
For a thin accretion disk surrounding an MBH of mass $M$, we adopt a Newtonian prescription for the mass accretion rate~\cite{Derdzinski:2020wlw,Zwick:2021dlg},
\begin{equation}
 \dot{M} \simeq 3\pi \nu \Sigma ,
\end{equation}
where $\nu$ is the kinematic viscosity of the disk and $\Sigma$ is its surface density. In the standard $\alpha$-disk model, the viscosity is parameterized in terms of a dimensionless constant $\alpha_{\rm disk}$ and the disk scale height $H(r)$ as
\begin{equation}
\nu = \alpha_{\rm disk}\, c_s\, H(r),
\end{equation}
with $\alpha_{\rm disk}\sim 0.001$-$0.1$~\cite{Davis:2009sc,Jiang:2019bxn}. The quantity $c_s$ is the isothermal sound speed, which is related to the Keplerian orbital frequency $\Omega_K$ through
\begin{equation}
c_s = H\,\Omega_K.
\end{equation}
To capture the main features of a thin accretion disk around the MBH in a simple parameterized form, we model the surface density profile as
\begin{eqnarray}
  \Sigma(r) = \Sigma_0 \left(\frac{r}{10M}\right)^{-\Sigma_p},
\end{eqnarray}
where  $\Sigma_0 \in[10^3,10^5]~\rm g/cm^3$,
and the aspect ratio $h(r)$ related to disk scale height $H(r)$ can be given by
\begin{eqnarray}
  h(r) = H(r)/r = h_0  \left(\frac{r}{10M}\right)^{0.5\Sigma_p-0.25}\;,
\end{eqnarray}
where $h_0\in[0.01, 0.1]$ and $\Sigma_0$ are constrained by the current astrophysical observation \cite{Jiang:2019ztr}.
In Ref.~\cite{Duque:2024mfw}, it is assumed that the relative velocity $\Delta v$ between the secondary and the local disk gas characterizes their gravitational interaction. For motion in the equatorial plane, the relative velocity between the secondary's orbital motion and the rotation of the disk can be written as
\begin{equation}
\Delta v \sim  \frac{e}{h} c_s.
\end{equation}
The secondary's orbital velocity is subsonic if the eccentricity is less than the disk's height, $e\leq h$. For the case of bigger eccentricity $e> h$, EMRI orbital's velocity is supersonic.
Note that if the disk's height parameter satisfies condition $h\geq0.02$ in the inner region, the secondary object's velocity is also supersonic for lower eccentric orbits.

In this work, we assume that the EMRI has a relatively large initial eccentricity and that the secondary moves supersonically through the accretion disk. In this regime, the disk exerts a drag (dynamical-friction) force on the secondary~\cite{Vicente:2019ilr,Muto:2011qv,Canto:2012bg},
\begin{equation}
F_{\rm DF}^{\rm disk} = - 2\pi \mu \,\frac{\Sigma}{2H}\,\frac{\eta}{\Delta v^2},
\end{equation}
where $\mu$ is the mass of the secondary. This force can be averaged over one orbital period; such an averaging scheme has been implemented in hydrodynamical simulations of EMRIs~\cite{Muto:2011qv}.

Following Ref.~\cite{Duque:2024mfw}, the relative velocity can be decomposed into radial and azimuthal components in cylindrical coordinates,
\begin{equation}
\Delta \mathbf{v} = \Delta v_r\, \mathbf{e}_r + \Delta v_\psi\, \mathbf{e}_\psi,
\end{equation}
with
\begin{eqnarray}
\Delta v_r &=& \frac{M}{p} \sin \phi,\\
\Delta v_\psi &=& \frac{M}{p} (1+\cos \phi) - \frac{M(1-e^2)}{p},
\end{eqnarray}
where $(\mathbf{e}_r, \mathbf{e}_\psi)$ denote the radial and azimuthal unit vectors, respectively, and $(p,\phi)$ are the semi-latus rectum and true anomaly of the Keplerian orbit.

Using the formalism of Ref.~\cite{Duque:2024mfw}, the orbit-averaged evolution of the EMRI orbital elements due to the accretion disk can be written as
\begin{equation}\label{eq:disk:pdot:edot}
\Bigg\langle\frac{d\mathcal{P}_i}{dt}\Bigg\rangle^{\rm disk}
= \frac{1}{T} \int_0^{2\pi} d\phi \left(\frac{d\mathcal{P}_i}{d\phi}\right),
\end{equation}
where $\mathcal{P}_i=\{\mathcal{P}_1,\mathcal{P}_2\}=\{p,e\}$ and
\begin{equation}
T=\int_0^{2\pi} d\phi \left(\frac{dt}{d\phi}\right)
\end{equation}
is the orbital period. The explicit expressions for $(dt/d\phi,dp/d\phi,de/d\phi)$ are given in Eqs.~(23)--(25) of Ref.~\cite{Duque:2024mfw}.

Throughout this paper, we assume that the secondary orbits within the inner region of an $\alpha$-disk, with orbital separation $r\lesssim 100M$ and $\Sigma_p=3/2$. When the true anomaly takes the values $\phi=0$ and $\phi=2\pi$, the secondary is at periapsis: there, dynamical friction slowly extracts orbital energy and the secondary is gradually decelerated by the surrounding gas. At apoapsis ($\phi=\pi$), on the other hand, the orbital velocity is increased by the interaction with the hot gas flow in the disk~\cite{Duque:2024mfw}.

\begin{figure*}[htb!]
\centering
\includegraphics[width=5.57in, height=3.2in]{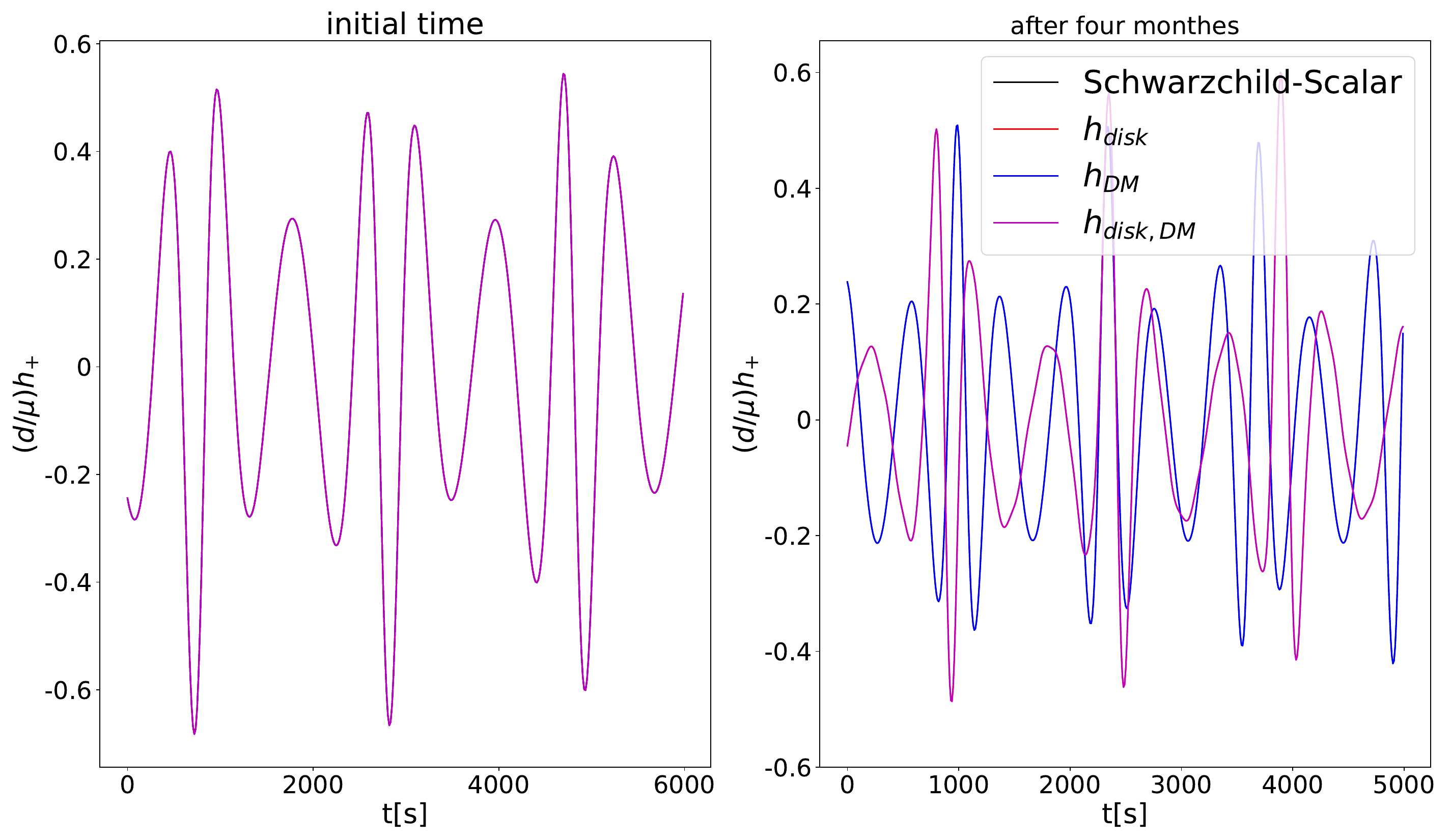}
\caption{Comparison of the plus polarization $h_{+}(t)$ for four EMRI waveforms in the time domain, illustrating the impact of different physical effects. The mass-ratio of binary objects is fixed to $\eta=10^{-5}$ and the initial orbital parameters are $(p_0,e_0)=(12,0.3)$. The environmental parameters are chosen as $\Sigma_p = 3/2$, $h_0 = 0.02$, $\Sigma_0 = 5.25\times10^3\,{\rm g\,cm^{-3}}$, $\alpha_{\rm DM}=1.8$, and scalar charge $q_s=0.1$. The black curve corresponds to the EMRI evolution including only the correction due to scalar emission. The red curve shows the waveform when the accretion-disk effect is added on top of the vacuum GR case. The cyan curve isolates the contribution from a DM minispike, while the purple curve represents the combined impact of both environmental effects, namely the accretion disk and the DM minispike.} \label{Fig:wave1}
\end{figure*}
\begin{figure*}[htb!]
\centering
\includegraphics[width=5.57in, height=3.2in]{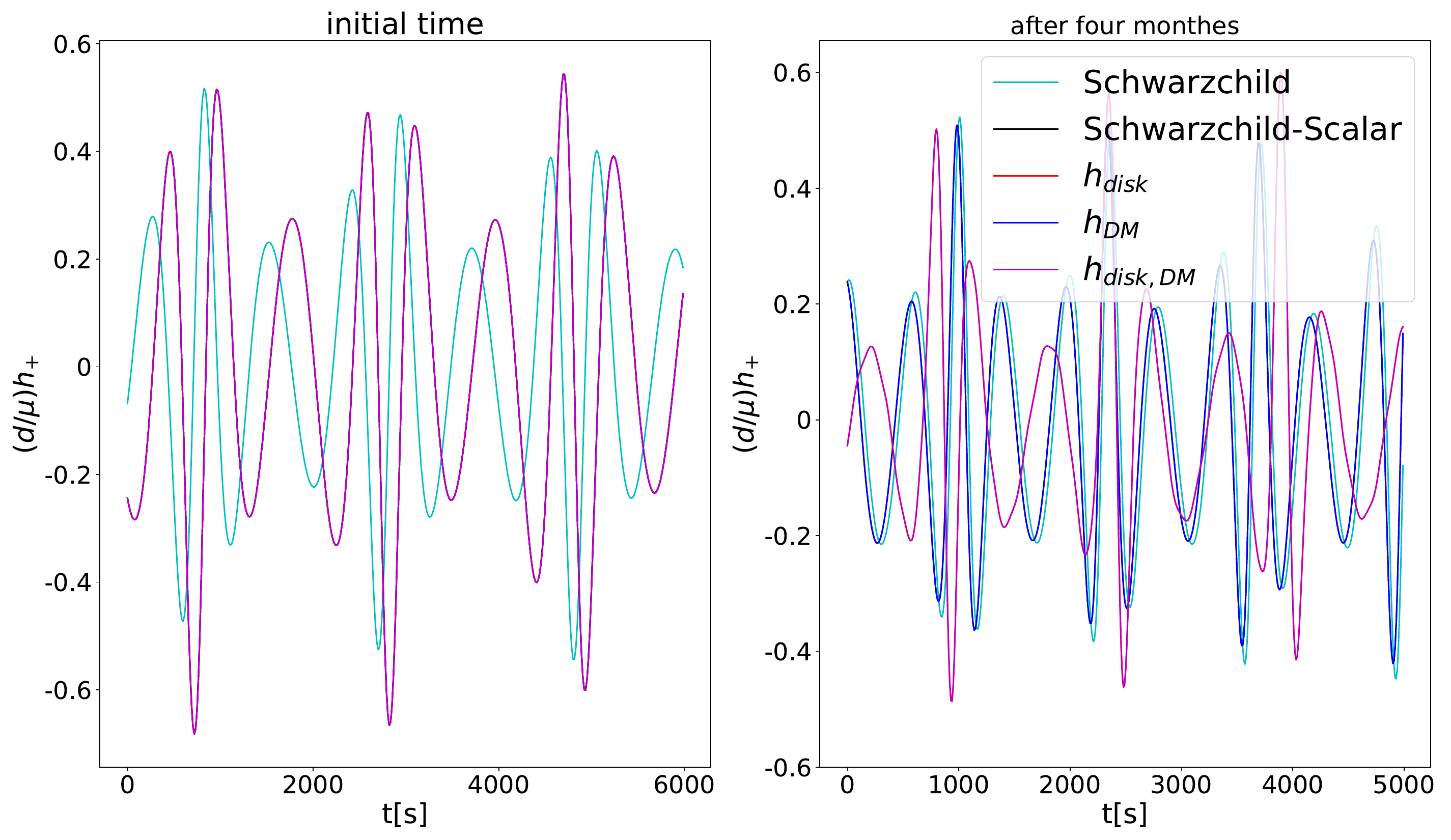}
\caption{Comparison of the plus polarization $h_{+}(t)$ for five EMRI waveforms in the time domain, highlighting the impact of different physical effects. The mass-ratio of binary objects is fixed to $\eta = 10^{-5}$, and the initial orbital parameters are $(p_0,e_0) = (12,0.3)$. The cyan curve shows the EMRI waveform in a vacuum Schwarzschild spacetime, while the remaining curves correspond to the same environmental configurations as in Fig.~\ref{Fig:wave1}.}

\label{Fig:wave2}
\end{figure*}

\begin{figure*}[htb!]
\centering
\includegraphics[width=3.17in, height=2.7in]{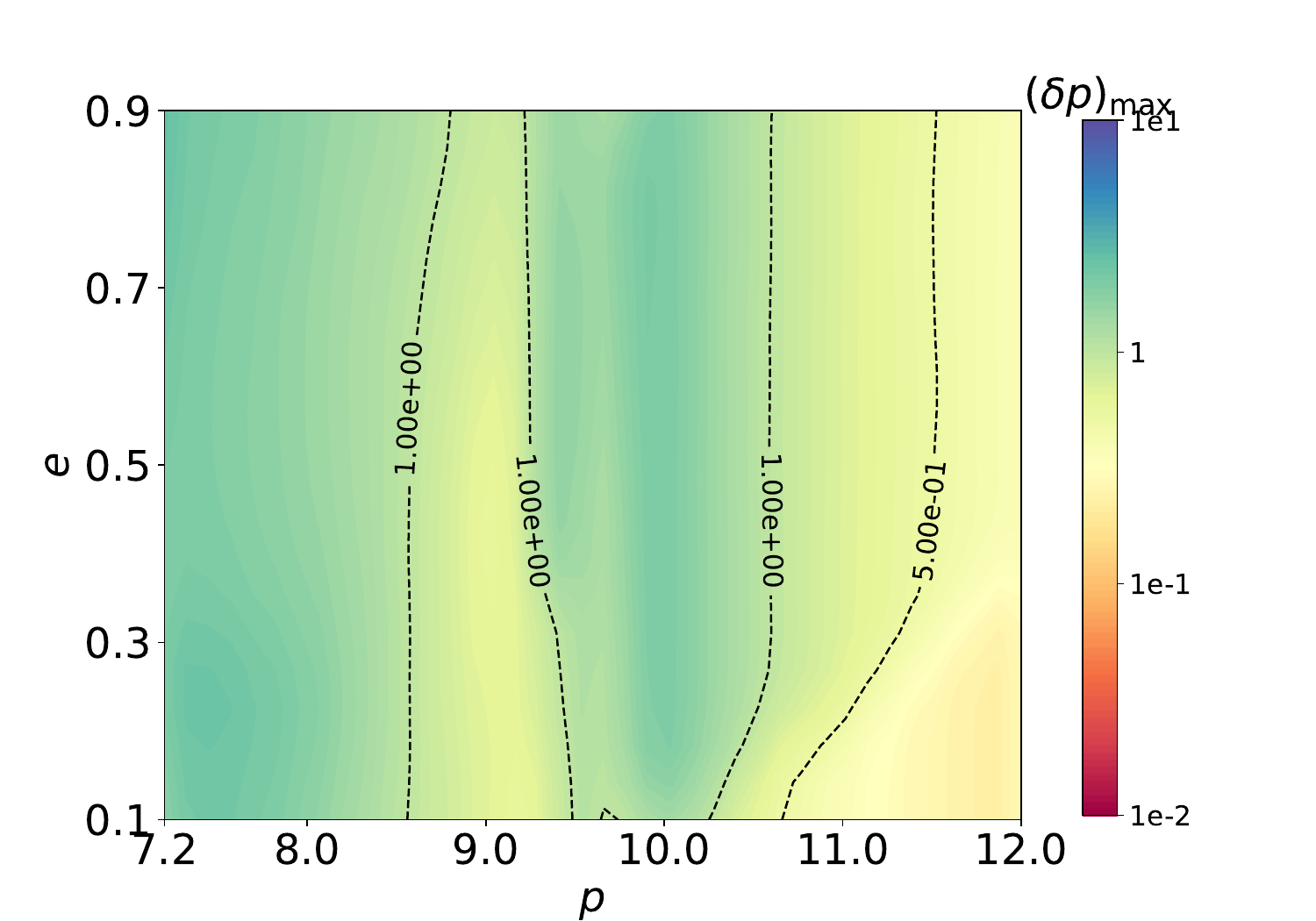}
\includegraphics[width=3.17in, height=2.7in]{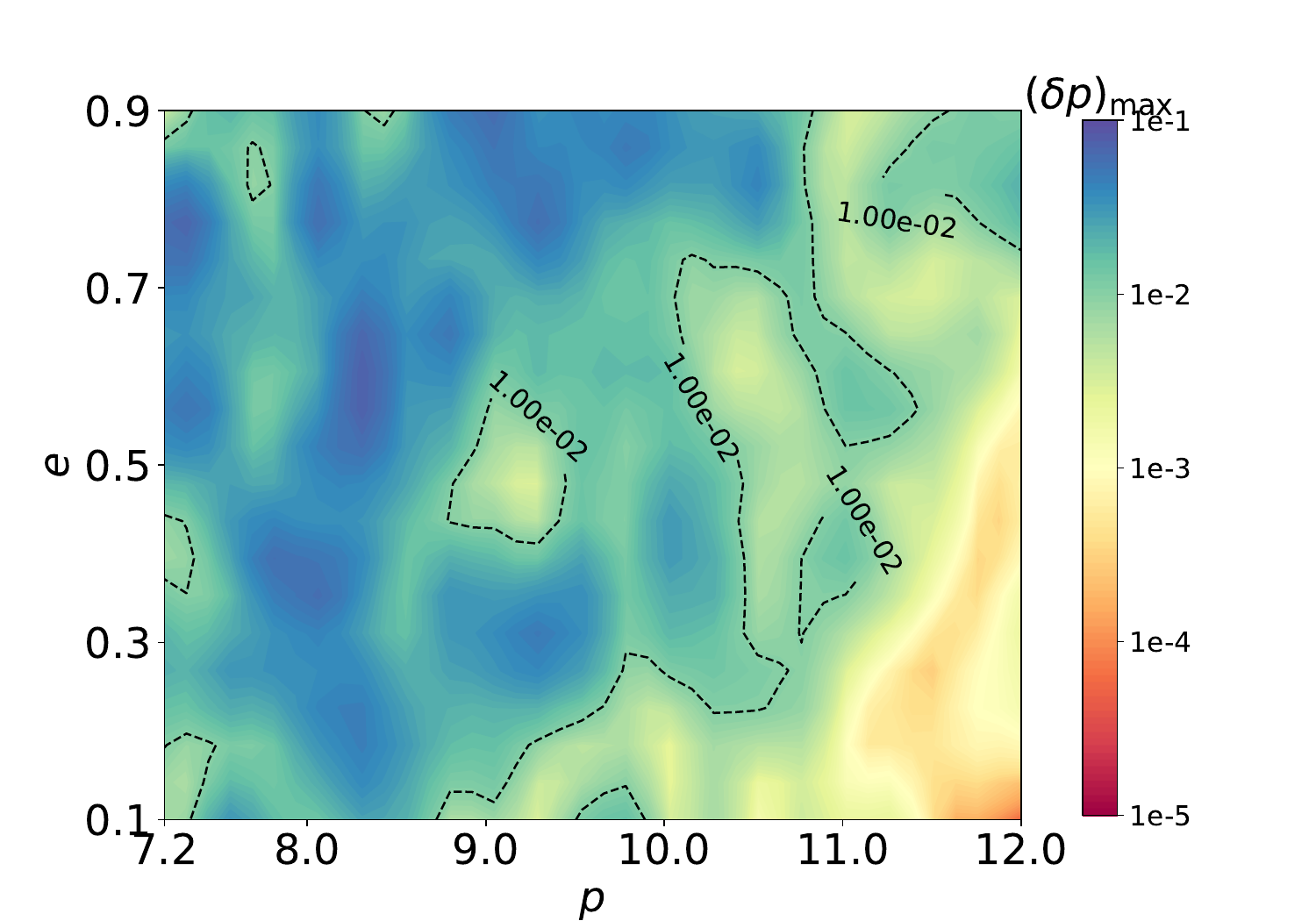}
\includegraphics[width=3.17in, height=2.7in]{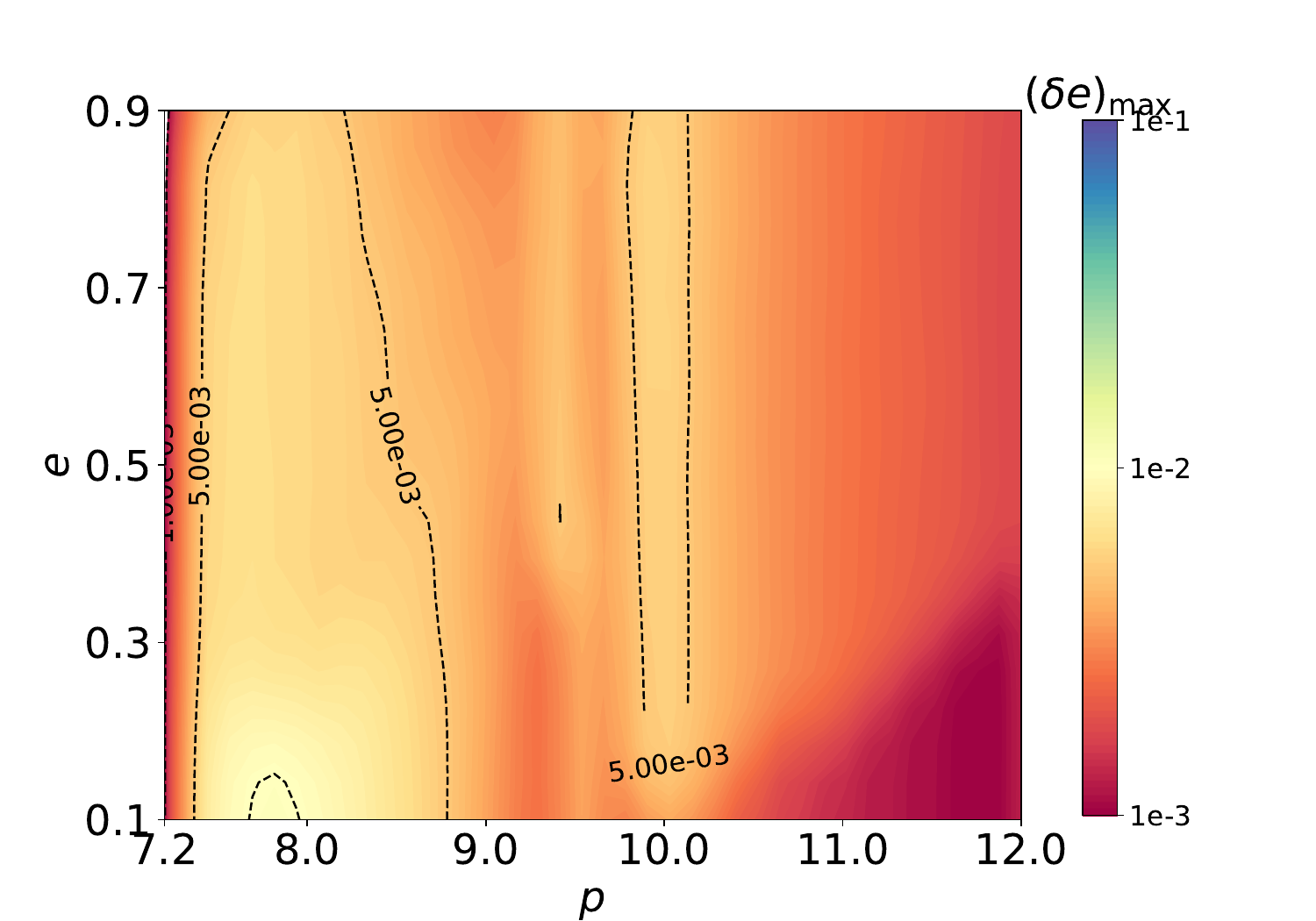}
\includegraphics[width=3.17in, height=2.7in]{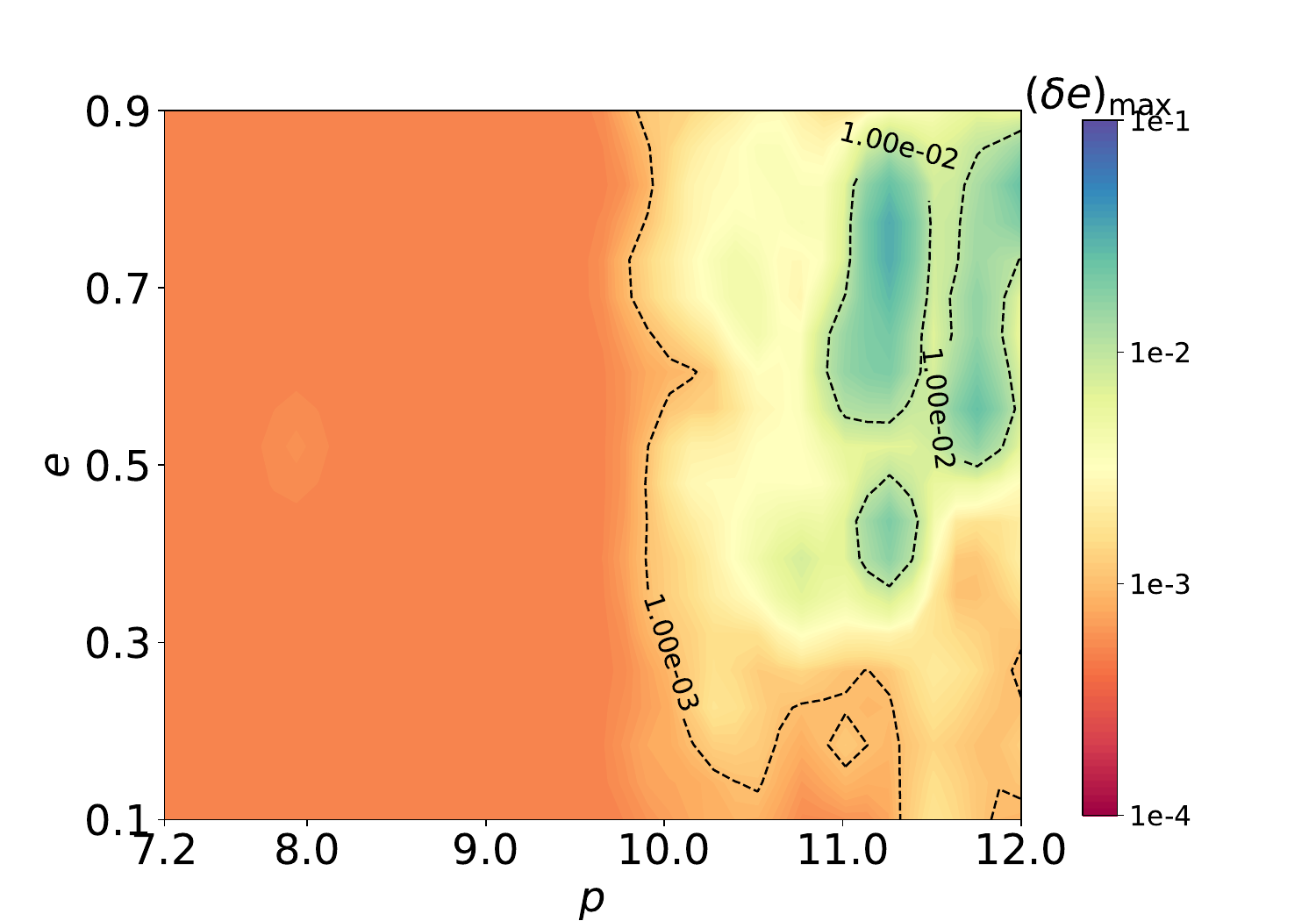}
\caption{Maximum deviations in the evolution of the orbital parameters $(p(t),e(t))$ induced by various environmental effects on EMRI dynamics, shown as functions of the orbital semi-latus rectum and eccentricity. The quantities $(\delta p)_{\rm max}$ and $(\delta e)_{\rm max}$ indicated above the color bars denote the maximum deviations in the evolution of $p(t)$ and $e(t)$ over the full duration of the eccentric inspirals. Black dashed curves indicate contours of constant maximum deviation in all four panels. The left panels display the maximum deviations between the vacuum spacetime and the accretion-disk case, while the right panels compare the vacuum spacetime with the DM environment. All remaining parameters associated with the environmental effects and the scalar charge are identical to those in Fig.~\ref{Fig:wave1}.}
\label{Fig:deltaP1}
\end{figure*}

\subsection{Dark matter halo}
The DM minispike around the MBH is modeled by a power-law density profile,
\begin{equation}
 \rho(r) = \rho_{\rm sp}\left(\frac{r_{\rm sp}}{r}\right)^{\alpha_{\rm DM}},
 \qquad r_{\rm ISCO} < r < r_{\rm sp},
\end{equation}
where $r_{\rm sp}$ is a characteristic scale radius related to the MBH influence radius $r_h$ by $r_{\rm sp} = 0.2\,r_h$, and $r_{\rm ISCO}$ is the radius of the innermost stable circular orbit. The influence radius $r_h$ is defined implicitly through
\begin{equation}
 M = 4\pi \int_0^{r_h} \rho_{\rm DM}\, r^2\, dr ,
\end{equation}
with $M$ the mass of the MBH. The parameter $\rho_{\rm sp}$ denotes the DM density at $r=r_{\rm sp}$; following Ref.~\cite{Eda:2014kra}, we adopt $r_{\rm sp}=0.54~\rm pc$ and $\rho_{\rm sp}=226\,M_\odot\,\rm pc^{-3}$. The power-law index $\alpha_{\rm DM}$ controls the slope of the minispike and is taken in the range $1.5\leq\alpha_{\rm DM}<2.3$~\cite{Eda:2013gg,Eda:2014kra}.

In this paper, we assume that the secondary object moves through the DM environment and experiences a dynamical-friction force~\cite{Chandrasekhar:1943ys,Eda:2014kra},
\begin{equation}
F_{\rm DF}^{\rm DM} = \frac{4\pi \mu^2  \rho_{\rm DM} \ln \Lambda}{v},
\end{equation}
where $\mu$ is the mass of the secondary, $v$ is its orbital velocity, and the Coulomb logarithm is taken to be $\ln \Lambda=20$~\cite{Amaro-Seoane:2007osp}. Within a Newtonian approximation~\cite{Yue:2019ozq}, the orbital speed can be written as
\begin{equation}\label{eq:dm:velocity}
v = \sqrt{\dot{r}^2 + r^2 \dot{\phi}^2}
= \left[-\frac{M(1-e^2)}{p} + \frac{2M(1+e\cos\psi)}{p}\right]^{1/2},
\end{equation}
where $\psi$ is the angular position of the secondary with respect to the MBH, and the final expression follows from the Keplerian relation $r = p/(1+e\cos\psi)$.

Thus, the orbit-averaged energy-loss rate due to dynamical friction in the DM minispike can be obtained as \cite{Yue:2019ozq}
\begin{eqnarray}
\Bigg<\frac{dE}{dt}\Bigg>^{\rm DM}_{\rm DF} &=&  \frac{1}{T} \int_0^T  F^{\rm DM}_{\rm DF} v dt
= (1-e^2)^{3/2}\int_0^{2\pi} d\psi
\nonumber \\ & \times & \frac{4\pi\mu^2 \rho_{\rm sp} r_{\rm sp}^{\alpha_{\rm DM}} \ln \Lambda (1+e\cos\psi)^{\alpha_{\rm DM} -2 } }{ p^{\alpha_{\rm DM}-1/2} M^{1/2} (1+2e\cos\psi + e^2)^{1/2} }
\end{eqnarray}
To concisely write the averaged fluxes, we omit the angle brackets in the following section.
On the other hand, the averaged angular momentum flux can be simplified the following form with the relation $<dL_z/dt> = r F_{\rm DF}^{\rm  DM} (r\dot{\psi}/v)$
\begin{eqnarray}
\Bigg<\frac{dL_z}{dt}\Bigg>^{\rm DM}_{\rm DF} &=& \frac{1}{T} \int_0^T  F^{\rm DM}_{\rm DF} v dt
= (1-e^2)^{3/2}\int_0^{2\pi} d\psi
\nonumber  \\ &\times & \frac{2\mu^2 \rho_{\rm sp} r_{\rm sp}^{\alpha_{\rm DM}} \ln \Lambda (1+e\cos\psi)^{\alpha_{\rm DM} -2 } }{ p^{\alpha_{\rm DM} -2} M^{1/2} (1+2e\cos\psi + e^2)^{3/2} }\;.
\end{eqnarray}
Here we model the effect of radiation reaction on the secondary by an effective friction force within a Newtonian framework, which we regard as a provisional prescription for the associated energy and angular-momentum fluxes and which will be upgraded to a fully relativistic treatment in future work~\cite{Cardoso:2022whc,Duque:2023seg}.
\subsection{Stellar-mass body with scalar hair in EMRIs and adiabatic evolution}
Previous studies have shown that ~\cite{Maselli:2020zgv,Maselli:2021men,Barsanti:2022vvl,Zi:2024lmt,Speri:2024qak}, within a certain class of scalar-tensor theories, the secondary in a typical EMRI can evade standard no-hair theorems and carry scalar hair sourced by the non-negligible curvature in the vicinity of the stellar-mass object. By contrast, the primary MBH can be approximately described by the no-hair theorem: the curvature in its surroundings is comparatively weaker than that near the stellar-mass companion, so any scalar hair on the MBH is negligible for our purposes.
In what follows, we therefore consider EMRIs in which a scalarized stellar-mass body orbits a Schwarzschild MBH, and we use the Teukolsky formalism in a non-rotating spacetime to compute the associated scalar radiation.

The equatorial geodesics of a test particle in Schwarzschild coordinates $(t,r,\theta,\phi)$ are characterized by the specific energy $E$ and angular momentum $L_z$, and satisfy
\begin{eqnarray}
\frac{dt}{d\tau} &=& \frac{E}{f(r)}\;, \qquad
\frac{d\phi}{d\tau} = \frac{L_z}{r^2}\;,\\[0.2cm]
\left(\frac{dr}{d\tau}\right)^2 &=& E^2 - V(r,L_z)\;,
\end{eqnarray}
where $f(r)=1-2M/r$, $\tau$ is the proper time, and
\begin{equation}
V(r,L_z) = f(r)\left(1+\frac{L_z^2}{r^2}\right)
\end{equation}
is the effective radial potential.
For bound eccentric motion, the turning points $r_p$ (pericenter) and $r_a$ (apocenter) satisfy $V(r_p,L_z)=V(r_a,L_z)=E^2$.
Introducing the semi-latus rectum $p$ and the eccentricity $e$ via the usual parametrization, one obtains the familiar analytic expressions
\begin{eqnarray}
E^2 &=& \frac{(p-2-2e)(p-2+2e)}{p\,(p-3-e^2)}\;, \quad
L_z = \sqrt{\frac{p^2 M^2}{p-3-e^2}}\;.
\end{eqnarray}

Following Ref.~\cite{Cutler:1994pb}, the geodesics can be parametrized by the relativistic anomaly $\chi$, defined through
\begin{equation}
r(\chi) = \frac{pM}{1+e\cos\chi}\;.
\end{equation}
The evolution of $t$ and $\phi$ with respect to $\chi$ then reads
\begin{eqnarray}
\frac{d\phi}{d\chi} &=& \sqrt{\frac{p}{p-6-2e\cos\chi}}\;, \\
\frac{dt}{d\chi} &=& \frac{p^2 M}{\bigl(p-2-2e\cos\chi\bigr)\bigl(1+e\cos\chi\bigr)^2}
\nonumber \\ && \times
\sqrt{\frac{(p-2-2e)(p-2+2e)}{p-6-2e\cos\chi}}\;.
\end{eqnarray}
Using these relations, the radial period $T_r$ and the accumulated azimuthal angle $\Delta\phi$ over one radial cycle can be expressed in terms of complete elliptic integrals~\cite{Wei:2025lva}:
\begin{eqnarray}\label{eq:sch:period}
\Delta \phi &=& \int_0^{2\pi} d\chi \,\frac{d\phi}{d\chi}
= \sqrt{\frac{16p}{p-6-2e}}\,
\mathcal{K}\!\left(\frac{4e}{6+2e-p}\right)\;, \\
T_r &=& \int_0^{2\pi} d\chi \,\frac{dt}{d\chi} =
C_1(p,e)\,\mathcal{K}\!\left(\frac{4e}{6+2e-p}\right)
\nonumber \\ &&
+\, C_2(p,e)\,\mathcal{E}\!\left(\frac{4e}{6+2e-p}\right)
\nonumber \\ &&
+\, C_3(p,e)\,\Pi\!\left(\frac{2e}{e-1},\frac{4e}{6+2e-p}\right)
\nonumber \\ &&
+\, C_4(p,e)\,\Pi\!\left(\frac{2e}{p-2e-2},\frac{4e}{6+2e-p}\right)\;,
\end{eqnarray}
where $\mathcal{K}$, $\mathcal{E}$ and $\Pi$ are the complete elliptic integrals of the first, second and third kind, respectively, and $C_{1,2,3,4}(p,e)$ are functions of the orbital parameters $(p,e)$.

The fundamental radial and azimuthal frequencies are then given by
\begin{equation}
\Omega_r = \frac{2\pi}{T_r}\;, \qquad
\Omega_\phi = \frac{\Delta \phi}{T_r}\;,
\end{equation}
and the generic orbital frequencies entering the multipolar decomposition are
\begin{equation}
\omega_{mk} = m\,\Omega_\phi + k\,\Omega_r\;,
\end{equation}
where $m$ and $k$ are integers.
The corresponding radial and azimuthal phases evolve according to
\begin{equation}
\frac{d\Phi_{r,\phi}}{dt} = \Omega_{r,\phi}\;,
\end{equation}
and we quantify the impact of environmental effects on the EMRI waveform through the dephasings
\begin{equation}\label{eq:delta:dephasing}
\delta\Psi_{r,\phi} = \Phi_{r,\phi}^{\rm vac} - \Phi_{r,\phi}^{\rm env}\;,
\end{equation}
where $\Phi_{r,\phi}^{\rm vac}$ denote the phases in vacuum and $\Phi_{r,\phi}^{\rm env}$ those in the presence of environmental effects.

In the weak-field approximation, one can derive the Teukolsky equations and their source terms analytically using a post-Newtonian (PN) expansion, and thereby obtain the scalar fluxes emitted by  EMRI~\cite{Zhang:2022rfr}.
For the scalar channel, the energy flux takes the form
\begin{eqnarray}
\frac{dE^{\rm scalar}}{dt}  &=& \frac{q_s^2 \mu}{3 M} v_s^8 \Bigg[1 - 2v_s^2 + 2\pi v_s^3 - 10v_s^4
+ \frac{12\pi v_s^5}{5}
\nonumber \\ &&
-\, e^2\left(1 - \frac{158 v_s^2}{15} - 3\pi v_s^3 + \frac{4268 v_s^4}{105} - \frac{47 \pi v_s^2}{3} \right)\Bigg]\;,
\end{eqnarray}
where $q_s$ is the scalar charge, $\mu$ is the scalar-field mass, and $v_s$ is the characteristic orbital velocity in the scalar sector.
The associated angular-momentum flux is related to the energy flux by
\begin{equation}
\frac{dL_z^{\rm scalar}}{dt}  = \frac{m}{\omega_{mk}}\,\frac{dE^{\rm scalar}}{dt}\;.
\end{equation}
The gravitational-wave fluxes $\left(\frac{dE^{\rm grav}}{dt},\,\frac{dL_z^{\rm grav}}{dt}\right)$ are computed using the analytic PN expansions of the energy and angular-momentum fluxes up to 19PN order~\cite{Munna:2022xts,Munna:2023vds}.
Since their explicit expressions are lengthy, we do not reproduce them here.

Including also the contribution from the DM-induced frictional force, the total fluxes in the non-vacuum are
\begin{eqnarray}
\frac{dE}{dt} &=& \frac{dE^{\rm grav}}{dt} + \frac{dE^{\rm scalar}}{dt} + \frac{dE^{\rm DM}}{dt}\;, \label{energy:Edot} \\
\frac{dL_z}{dt} &=& \frac{dL_z^{\rm grav}}{dt} + \frac{dL_z^{\rm scalar}}{dt} + \frac{dL_z^{\rm DM}}{dt}\;.\label{angular:Ldot}
\end{eqnarray}
These fluxes, arising from gravitational radiation, scalar emission, and environmental effects (DM and the accretion disk), drive the long-term orbital evolution of the EMRI.
Within the adiabatic approximation, the inspiral is treated as a sequence of geodesics parametrized by the integrals of motion $\mathcal{C}=(E,L_z)$, which evolve according to the balance law
\begin{equation}
\dot{\mathcal{C}} = -\,\dot{\mathcal{C}}_{\rm GW}\;,
\end{equation}
where an overdot denotes differentiation with respect to coordinate time.

The evolution of the orbital elements $(p(t),e(t))$ can then be obtained from the fluxes in Eqs.~\eqref{energy:Edot} and \eqref{angular:Ldot}.
Defining the Jacobian
\begin{equation}
H = \frac{\partial E}{\partial p}\frac{\partial L_z}{\partial e}
 - \frac{\partial E}{\partial e}\frac{\partial L_z}{\partial p}\;,
\end{equation}
we write
\begin{eqnarray}
\dot{p}_{\rm DM,scalar} &=&
\frac{\dot{E}\,\frac{\partial L_z}{\partial e}
      + \dot{L}_z\,\frac{\partial E}{\partial e}}{H}\;, \\
\dot{e}_{\rm DM,scalar} &=&
\frac{\dot{L}_z\,\frac{\partial E}{\partial p}
      - \dot{E}\,\frac{\partial L_z}{\partial p}}{H}\;,
\end{eqnarray}
where $\dot{E}$ and $\dot{L}_z$ include the gravitational, scalar, and DM friction contributions.
The additional dynamical-friction terms arising from the accretion disk, $(\dot{p}_{\rm disk},\dot{e}_{\rm disk})$, are computed from Eqs.~\eqref{eq:disk:pdot:edot}.
Altogether, the evolution of the orbital elements in the presence of environmental effects is
\begin{equation}\label{eq:pedot:env}
\dot{p}_{\rm env} = \dot{p}_{\rm DM,scalar} + \dot{p}_{\rm disk}\;, \qquad
\dot{e}_{\rm env} = \dot{e}_{\rm DM,scalar} + \dot{e}_{\rm disk}\;.
\end{equation}
Here, the subscript ``env'' denotes the combined influence of the DM minispike, the scalar radiation, and the accretion-disk friction; ``disk'' refers specifically to the accretion-disk contribution; and ``DM, scalar'' contains the effects of DM friction and scalar emission.

\begin{figure*}[htb!]
\centering
\includegraphics[width=3.17in, height=2.7in]{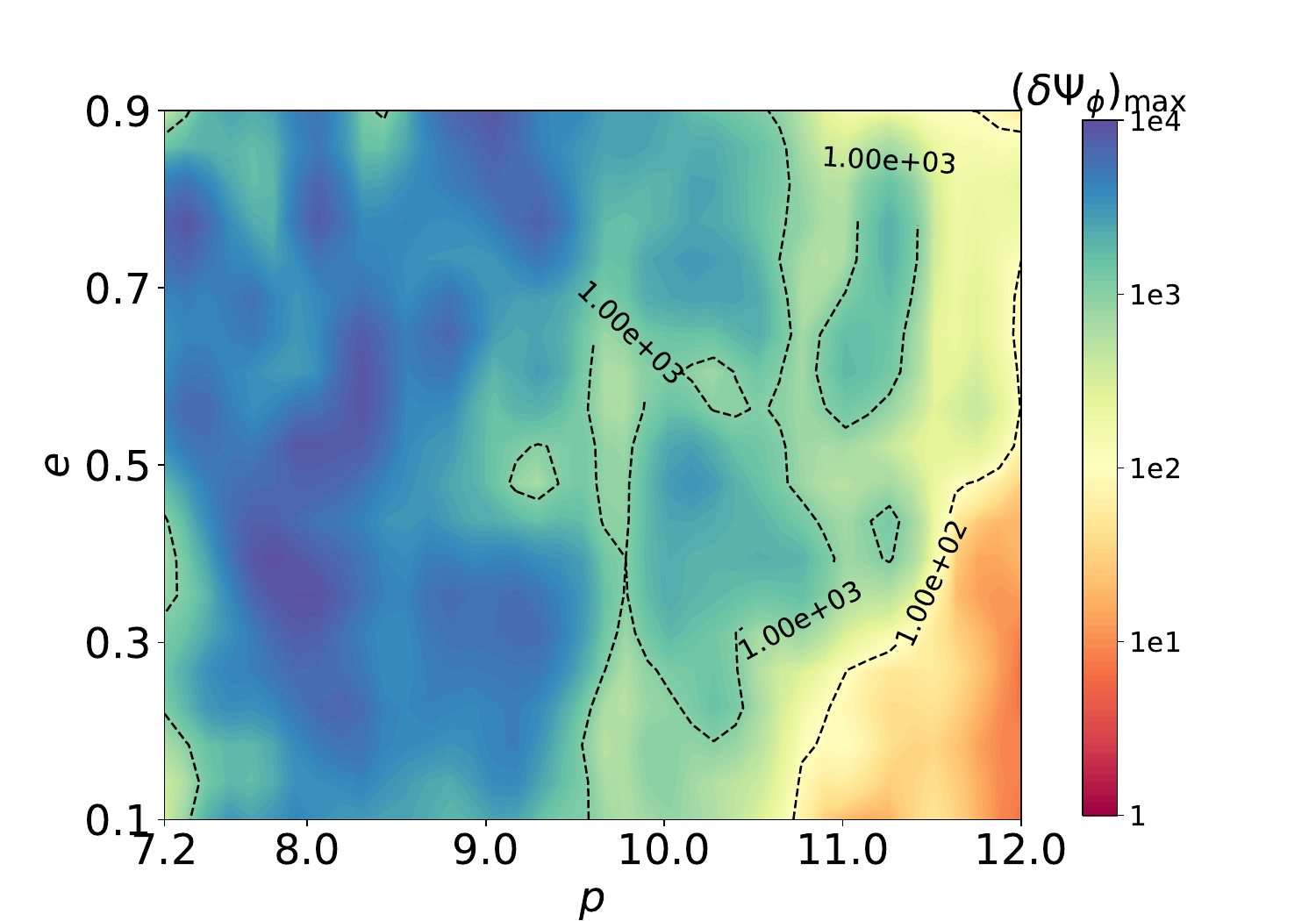}
\includegraphics[width=3.17in, height=2.7in]{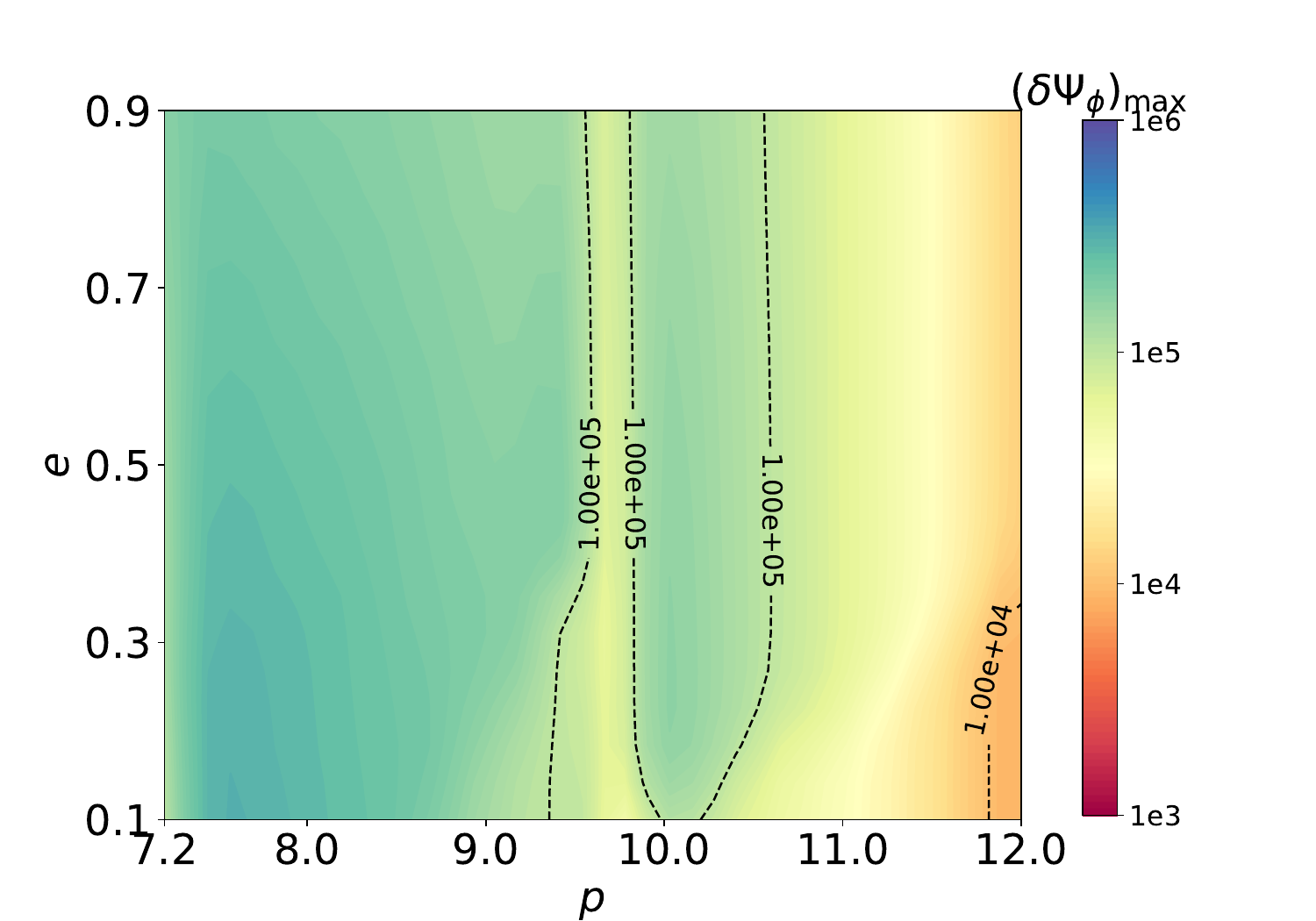}
\includegraphics[width=3.17in, height=2.7in]{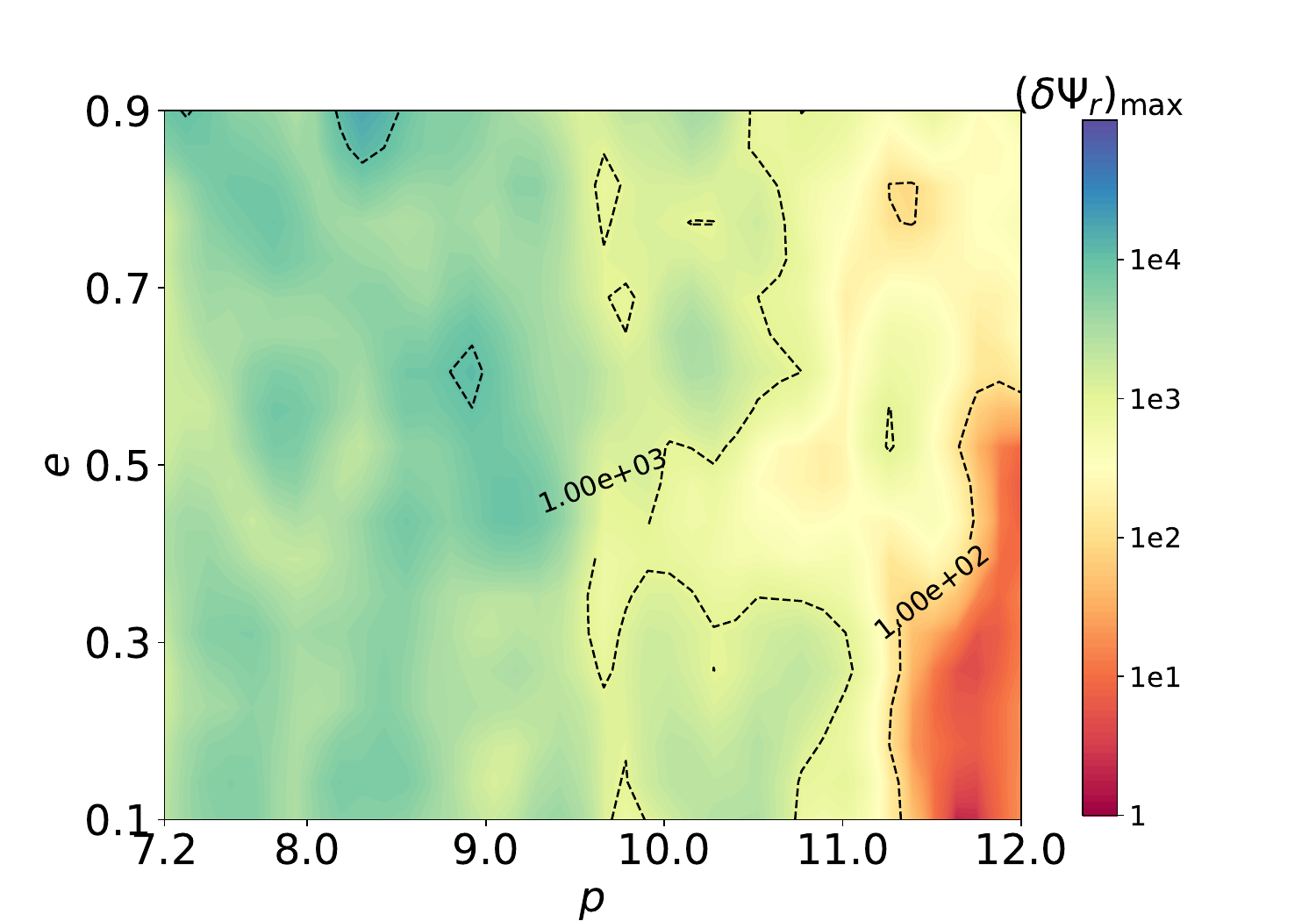}
\includegraphics[width=3.17in, height=2.7in]{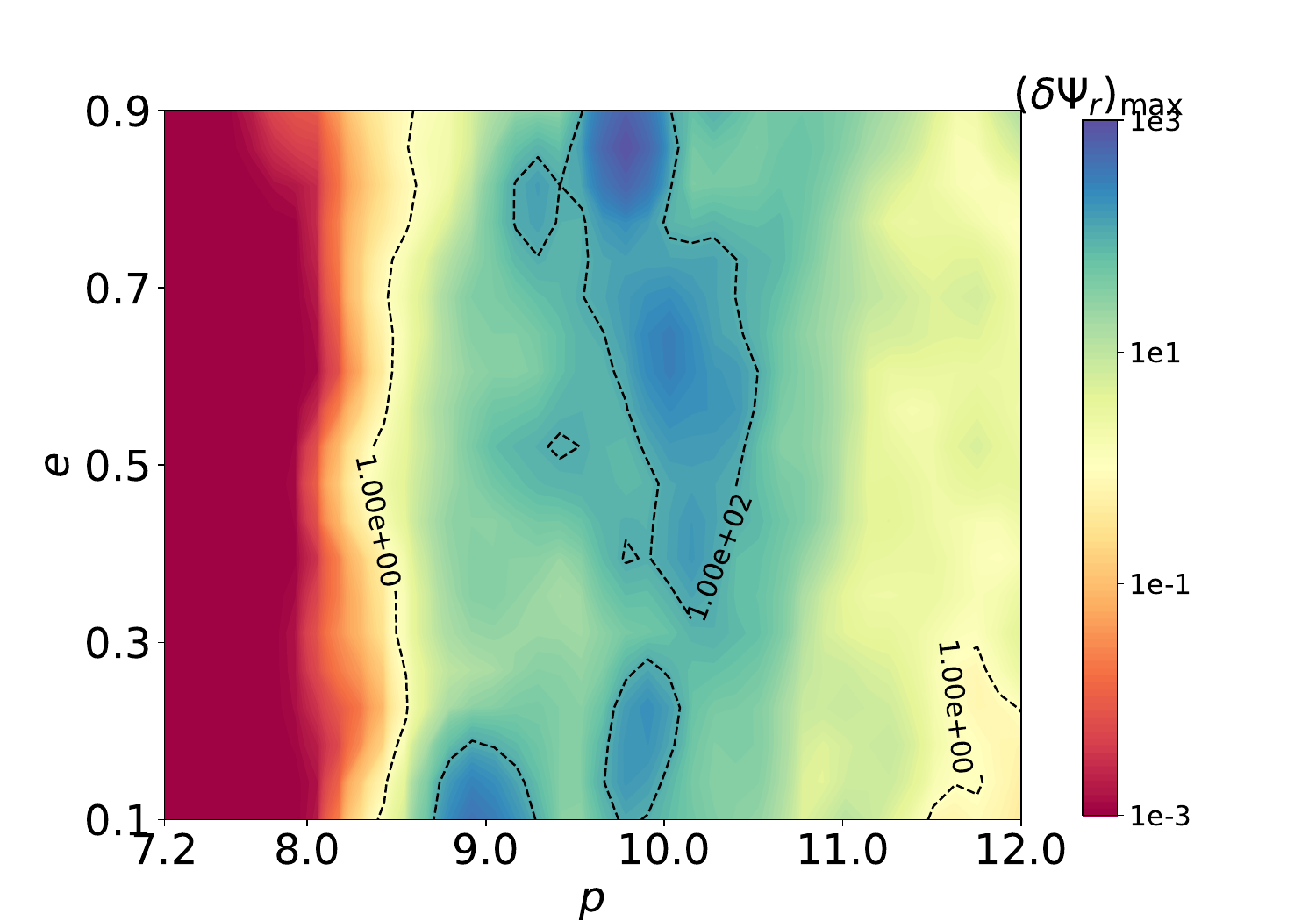}
\caption{Azimuthal and radial dephasing are shown as functions of the orbital semi-latus rectum and eccentricity, comparing scalar emission in vacuum with that in nonvacuum environments including dark-matter dynamical friction and an accretion disk. All remaining EMRI, environmental, and scalar-charge parameters are identical to those adopted in Fig.~\ref{Fig:wave1}.} \label{Fig:dephasing}
\end{figure*}

\begin{figure*}[htb!]
\centering
\includegraphics[width=3.17in, height=2.7in]{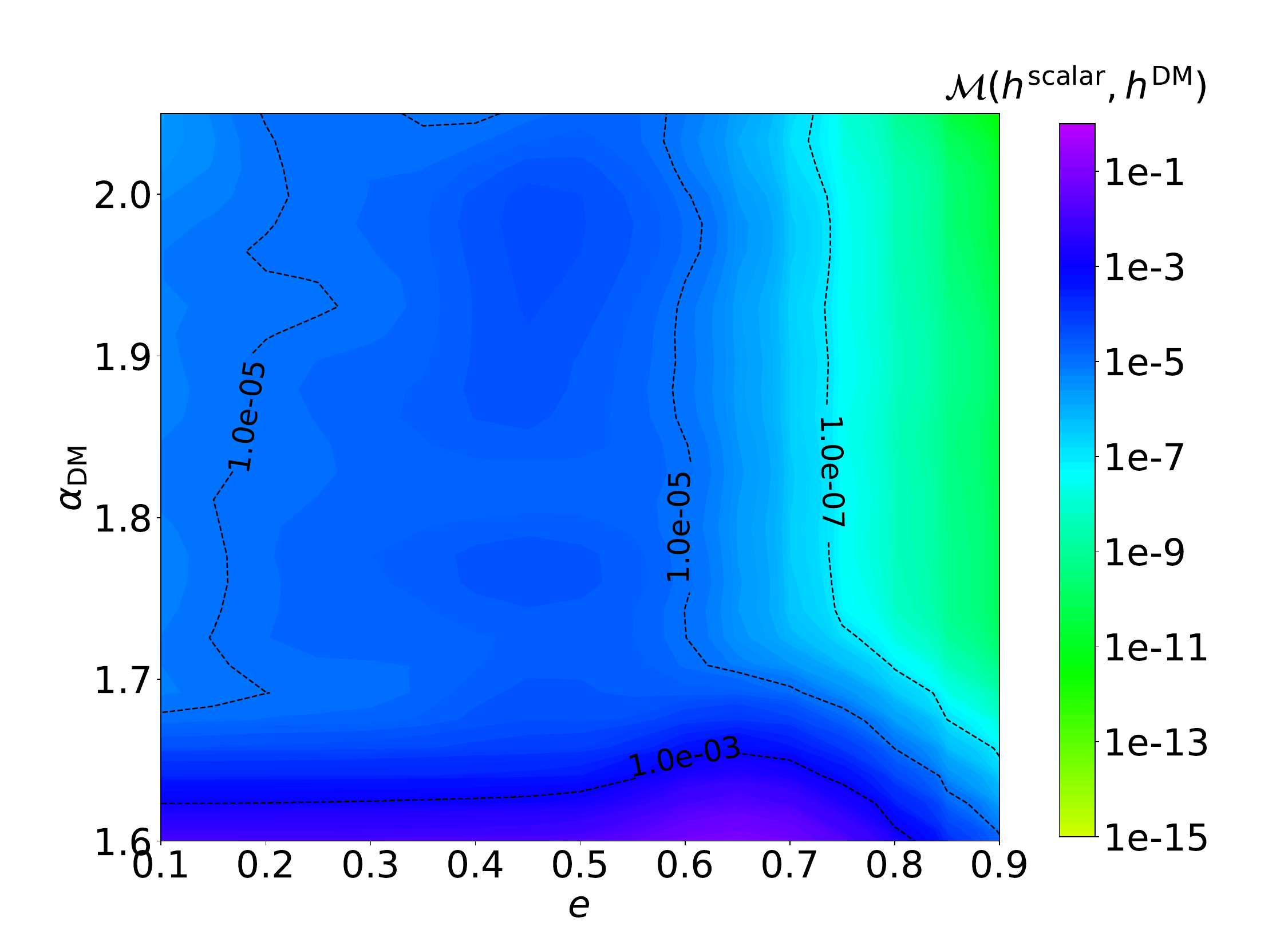}
\includegraphics[width=3.17in, height=2.7in]{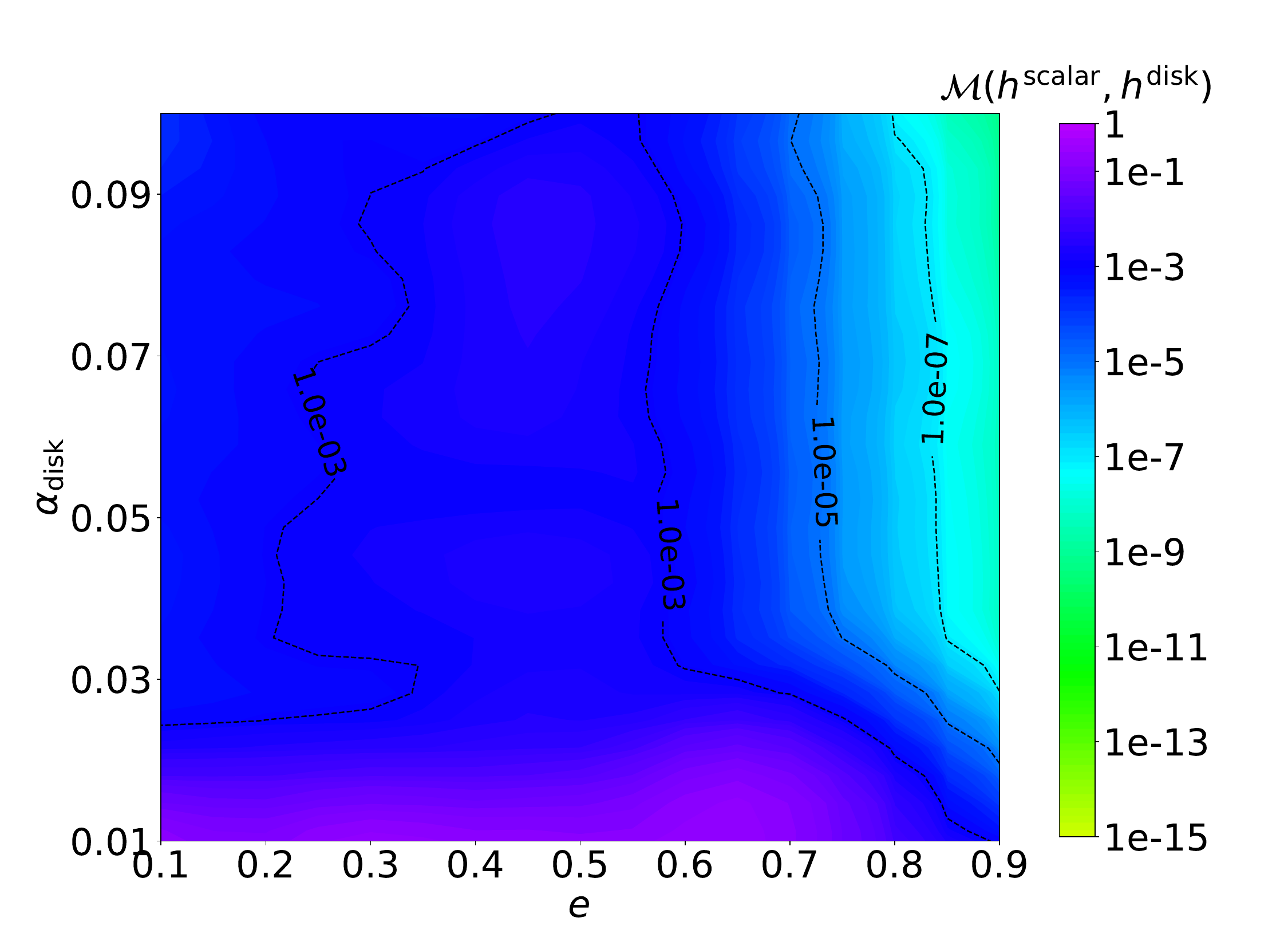}
\includegraphics[width=3.17in, height=2.7in]{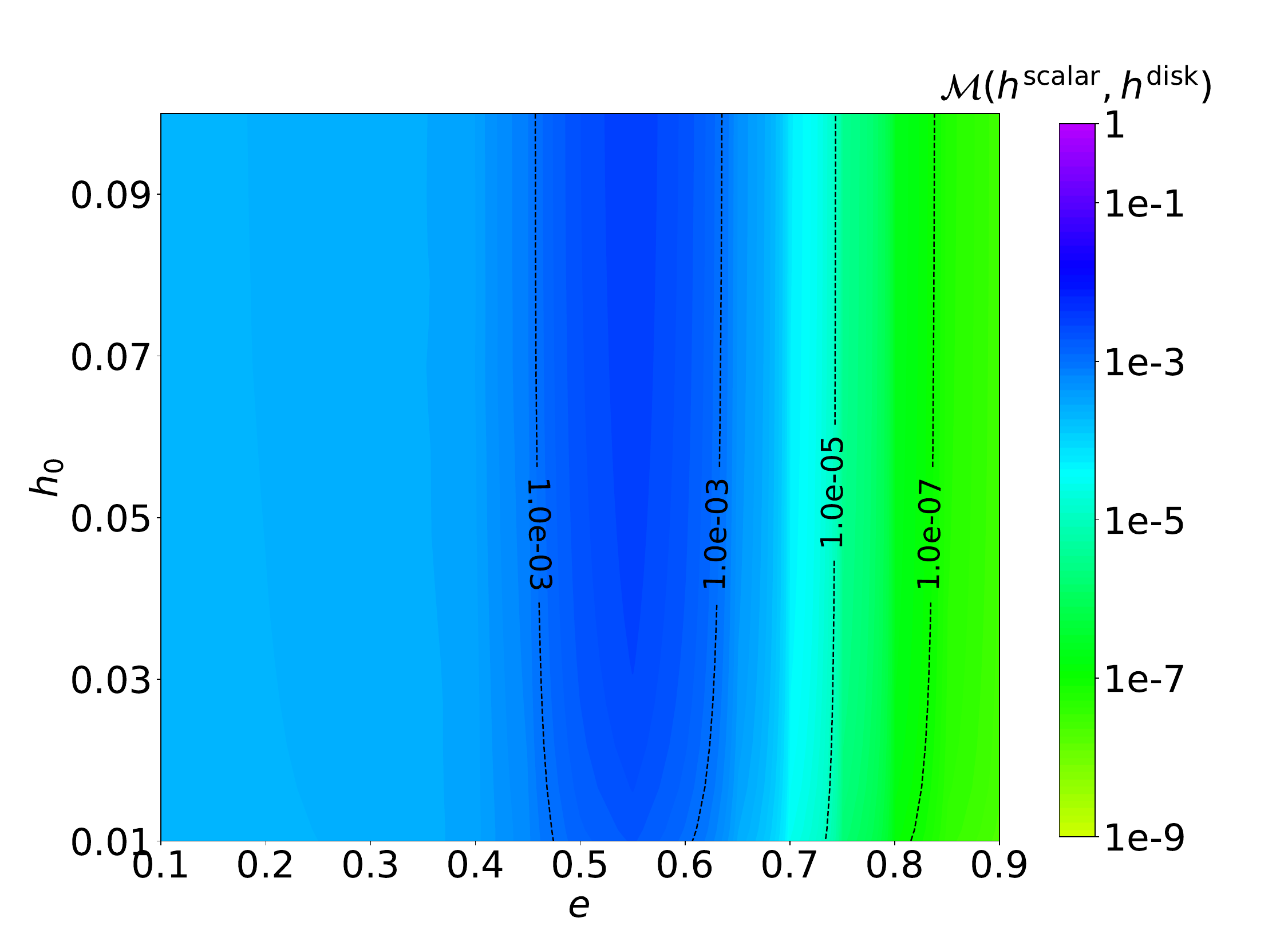}
\includegraphics[width=3.17in, height=2.7in]{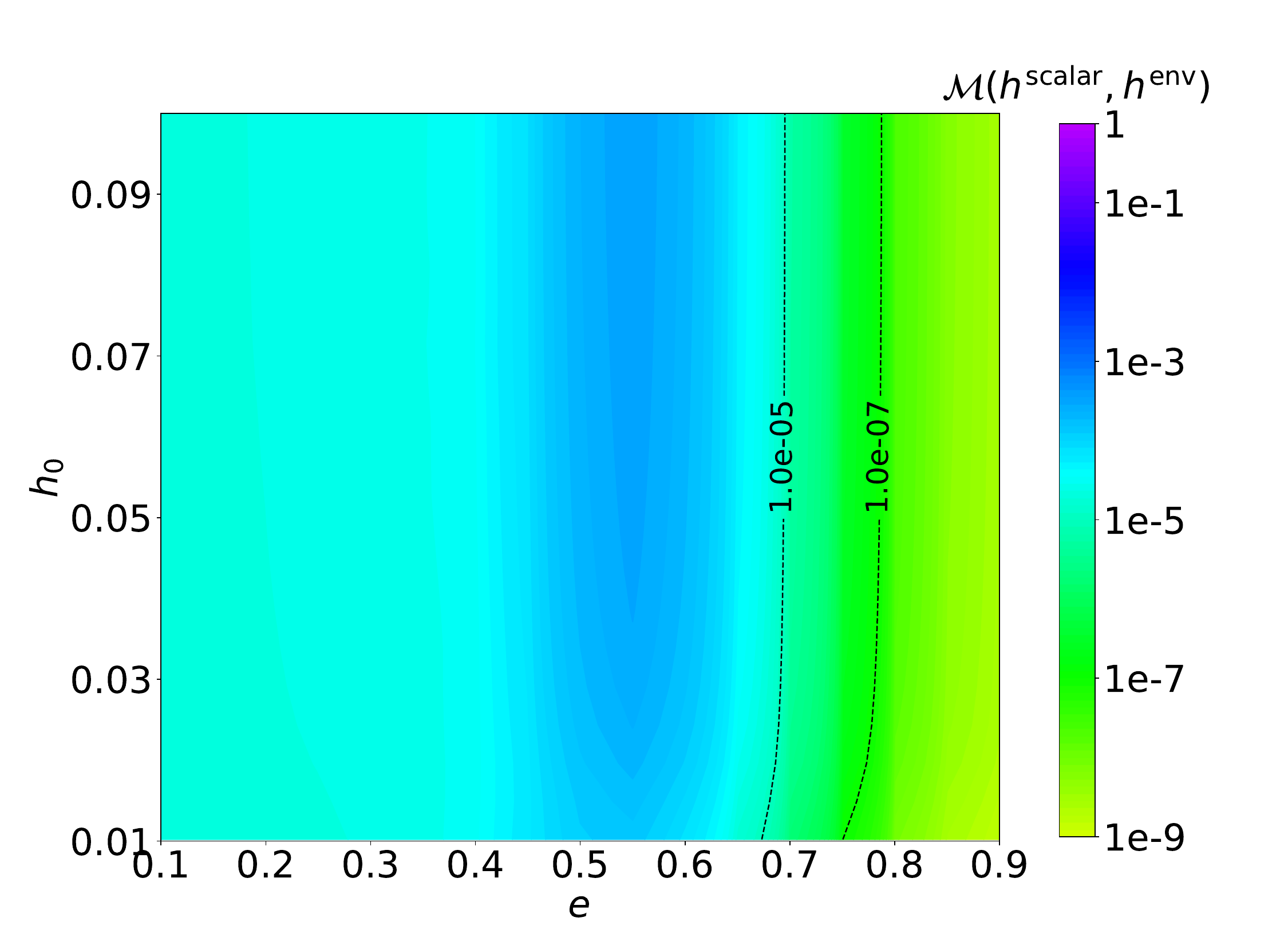}
\caption{Mismatches between EMRI signals with scalar emission and those including environmental effects from DM dynamical friction or accretion-disk interactions are shown as functions of the orbital eccentricity $e$ and the relevant environmental parameters. Four representative cases are considered with initial semi-latus rectum $p_0 = 12$, scalar charge $q = 0.01$, and all remaining parameters identical to those in Fig.~\ref{Fig:wave1}. The labels $h^{\rm scalar}$ and $h^{\rm DM}$ above the color bars denote EMRI waveforms corrected by scalar emission and by DM friction, respectively. The waveform $h^{\rm disk}$ corresponds to an EMRI embedded in an accretion disk, while $h^{\rm env}$ describes an EMRI evolving in a combined environment consisting of both a DM minispike and an accretion disk. The black dashed curves indicate contours of constant mismatch; in particular, the contour at $\mathcal{M} \simeq 10^{-3}$ marks the typical threshold for distinguishability adopted for LISA-like detectors. The remaining environmental parameters are chosen as $\alpha_{\rm disk}=0.04$, $h_0=0.05$ (top left panel), $\alpha_{\rm DM}=1.65$, $h_0=0.05$ (top right panel), $\alpha_{\rm DM}=1.65$, $\alpha_{\rm disk}=0.04$ (bottom left panel), and $\alpha_{\rm disk}=0.04$, $\alpha_{\rm DM}=1.65$ (bottom right panel).} \label{Fig:mismatch}
\end{figure*}

\subsection{Waveform and data analysis}\label{wave:data:analysis}
After computing the inspiraling trajectories using the hybrid fluxes in Eqs.~\eqref{eq:pedot:env}, we can obtain the EMRIs signal by LISA in the low-frequency approximate~\cite{Cutler:1997ta}.
Strictly speaking, we need to compute the fully relativistic EMRIs waveform \cite{Hughes:2021exa}. However, it is rather expensive and time-consuming to compute the adiabatic waveforms via multivoice decomposition.
Instead, we incorporate environmental effects into the waveform model by modifying the \texttt{Augmented Analytical Kludge (AAK)} framework~\cite{Chua:2017ujo}, feeding in inspiral trajectories that have been corrected for the various environmental contributions.

The \texttt{AAK} model is essentially based on a quadrupole-order approximation to the relativistic waveform, which substantially reduces the computational cost while retaining the key secular features of the signal.
Within this framework, two GW polarizations $h_+(t)$ and $h_\times(t)$ are obtained by summing over harmonics of the fundamental orbital frequencies, which can be given by
\begin{equation}
\begin{aligned}\label{amplitude}
h_+ = \sum_n &-\Big[1+(\hat{L}\cdot\hat{n})^2\Big]\Big[a_n \cos2\gamma -b_n\sin2\gamma\Big] \\
& +c_n\Big[1-(\hat{L}\cdot\hat{n})^2\Big], \\
h_\times  = \sum_n &2(\hat{L}\cdot\hat{n})\Big[b_n\cos2\gamma+a_n\sin2\gamma\Big],
\end{aligned}
\end{equation}
where $\hat{n}$ is the unit direction vector and $\hat{L}$ is the
unit vector of the orbital angular momentum. The coefficients $(a_n, b_n, c_n)$ are written in term of the eccentricity $e$ and mean anomaly $\Phi \equiv \Phi_r$, which have been derived \cite{Peters:1963ux}
\begin{equation}
\begin{aligned}
a_n =~ &-n \mathcal{A} \Big[J_{n-2}(ne)-2eJ_{n-1}(ne)+\frac{2}{n}J_n(ne)
\nonumber \\ &+ 2J_{n+1}(ne) -J_{n+2}(ne)\Big]\cos(n\Phi_r), \\
b_n =~ &-n \mathcal{A}(1-e^2)^{1/2}\Big[J_{n-2}(ne)-2J_{n}(ne)+J_{n+2}(ne)\Big]
\nonumber \\ & \times\sin(n\Phi_r), \\
c_n =~& 2\mathcal{A}J_n(ne)\cos(n\Phi_r), \\
\end{aligned}
\end{equation}
with a quantity $\mathcal{A} = ~ (2\pi M \Omega_\phi )^{2/3}\mu/d$ and the distance $d$ from source to detector.
Note that $J_n(ne)$ is the first kind Bessel functions relating to the eccentricity and $\gamma = \Phi_\phi - \Phi_r$ is the azimuthal angular that means the direction of eccentric orbital pericenter.

In order to assess the impact of environmental effects on the GW phase of EMRIs, we examine the differences in orbital parameters and phases obtained from Eqs.~\eqref{eq:delta:dephasing} and \eqref{eq:pedot:env}.
We first compare a set of representative EMRI waveforms in the time domain, including various environmental contributions, as shown in Figs.~\ref{Fig:wave1} and \ref{Fig:wave2}.
In each figure, the left panel displays the waveforms over the initial stage of the inspiral, while the right panel shows the corresponding signals after approximately four months of evolution.
At early times, the phases of the waveforms for different environmental configurations remain nearly indistinguishable; however, as the system evolves, the accumulated phase differences become clearly visible by eye in Fig.~\ref{Fig:wave1}.
In particular, in Fig.~\ref{Fig:wave2} the waveform generated in vacuum exhibits a substantial dephasing relative to those that include additional physical effects.

To quantify the impact of different environmental effects on the EMRI waveform in the presence of a scalar charge carried by the secondary, we compute the mismatch between pairs of signals corresponding to scalar emission and to DM friction or accretion-disk interactions.
The mismatch provides a standard measure of the distinguishability of two GW signals in a detector, and is defined as~\cite{Cutler:1994ys}
\begin{align}
\mathcal{M}(h_a,h_b) =& 1 - \mathcal{O}(h_a,h_b) \;, \label{mismatch}\\
\mathcal{O}(h_a, h_b) =& \frac{(h_{a}\vert h_{b})}{\sqrt{(h_{a}\vert h_{a})(h_{b}\vert h_{b})}},
\end{align}
where the inner product between two data streams
in the frequency domain are given by
\begin{equation}
(h_a|h_b) = 2 \int^{f_{\rm high}}_{f_{\rm low}} \frac{h_a^*(f)h_b(f)+h_a(f)h_b^*(f)}{S_n(f)}df.
\end{equation}\label{inner}
with $f_{\rm low}= 0.1 ~\rm m Hz$ and $f_{\rm high}$ corresponds to the orbital frequency around last stability orbit for the Schwarzchild spacetime. The function $S_{n}(f)$ is the power spectral density of LISA-like detector \cite{LISA:2017pwj}.
To assess LISA's ability to distinguish two GW signals, an empirical criterion is adopted: $\mathcal{M} \geq 1/(2\rho^2)$, where $\rho$ is the signal-to-noise ratio (SNR). LISA can resolve two signals if their mismatch satisfies this inequality.
According to the previous studies on assessing the detectable of EMRIs system, a typical SNR of the signal by LISA-like detectors is $\rho=20$ \cite{Babak:2017tow, Fan:2020zhy}. Consequently, the threshold of mismatch is $\mathcal{M}\sim 0.001$, serving as a benchmark of distinguishing the environmental effects from the scalar imprinting for EMRIs waveforms.

To evaluate the constraining modification of different environmental effects on EMRIs waveform with LISA, we show the measurement errors and their relevance  over the different parameters by computing the Fisher information matrix.
For a EMRI signal with a higher SNR, the uncertainties of source parameters $\boldsymbol{\theta}$ describing binaries can be approximately obtained by covariance matrix \cite{Vallisneri:2007ev}
\begin{eqnarray}
\sigma_{\boldsymbol{\theta}_i} =
\sqrt{\boldsymbol{\Sigma}_{\boldsymbol{\theta}_i \boldsymbol{\theta}_i}} \label{sigma:fim}\;,
\end{eqnarray}
where the source parametric vector $\boldsymbol{\theta}_i=\{M,\mu,p_0,e_0,\alpha_{\text{DM}},h_0,\Sigma_0,\Phi_{r,0},
\Phi_{\phi,0},d_L,\theta_S,\phi_S,\theta_K,\phi_K\}$ with $i=1,2,\dots,14$,
consists of the following parameters:
$M$ and $\mu$ denote the masses of the MBH and the secondary object;
$p_0$ and $e_0$ are the initial orbital semi-latus rectum and eccentricity; the parameters $\alpha_{\mathrm{DM}}, h_0$ and $\Sigma_0$ characterize the environmental effects;
$\Phi_{r, 0}$ and $\Phi_{\phi, 0}$ are the initial radial and azimuthal orbital phases;
$d_L, \theta_S, \phi_S$ specify the luminosity distance and sky location of the source;
and $\theta_K, \phi_K$ describe the polar and azimuthal angles of the MBH spin orientation.

The covariance matrix is given by the inverse of FIM,
\begin{equation}
  \Sigma_{ij}
  \equiv \big\langle \Delta\theta_i \,\Delta\theta_j \big\rangle
  = \big(\boldsymbol{\Gamma}^{-1}\big)_{ij} \, ,
\end{equation}
where the FIM is defined in terms of the GW signal and the source parameters as
\begin{equation}
  \Gamma_{ij}
  = \left(
      \frac{\partial h(f)}{\partial \theta_i}
      \Bigg|
      \frac{\partial h(f)}{\partial \theta_j}
    \right) ,
\end{equation}
and $(\,\cdot\,|\,\cdot\,)$ denotes the inner product weighted by the noise power spectral density of LISA-like detectors~\cite{LISA:2017pwj,TianQin:2015yph,TianQin:2020hid}.
The parameter uncertainties inferred from Eq.~\eqref{sigma:fim} therefore provide tighter constraints on the environmental parameters than the dephasing and mismatch criteria of Eqs.~\eqref{eq:delta:dephasing} and~\eqref{mismatch}.
Moreover, the off-diagonal elements of the inverse FIM encode the correlations among the source parameters, and thus quantify how the uncertainty in the scalar charge is affected by degeneracies with the other parameters.

\section{Result}\label{result}
In this section, we present the results of difference of phases and orbital parameter evolution to evaluate the environmental effects on EMRIs waveform.

In Fig.~\ref{Fig:deltaP1}, we show the differences in inspiraling trajectories obtained in the vacuum spacetime and in the presence of various environmental effects.
For a range of initial orbital parameters $(p,e)$, the upper panels show the maximum deviation $(\delta p)_{\rm max}$ of the semi-latus rectum for eccentric EMRI inspirals between the vacuum and non-vacuum cases over the long-duration evolution; the lower panels display the corresponding maximum deviation $(\delta e)_{\rm max}$ of the eccentricity between the vacuum and non-vacuum cases for the long-term evolution of the orbital eccentricity.
The right panels present a comparison of EMRIs orbits in a DM minispike with those in the vacuum spacetime, whereas the left panels consider the discrepancies of EMRI's orbits for a secondary object endowed with scalar charge in the vacuum and in the presence of the two environmental effects.
The black dashed lines represent contours of the trajectory differences in the various astrophysical environments. From the four panels in Fig.~\ref{Fig:deltaP1}, one can see that the environmental effects surrounding the MBH play a more significant role in the orbital evolution compared to the case with only the DM effect.
Therefore, the environmental effects indeed generate a larger deviation of the orbital-parameter evolution from the vacuum spacetime, which would be a distinguishable effect for LISA-like detectors.

Figure~\ref{Fig:mismatch} presents the mismatch $\mathcal{M}$ as a function of the orbital eccentricity $e$ and various environmental parameters, including $\alpha_{\rm DM}$, $\alpha_{\rm disk}$, and $h_0$. These results assess the distinguishability of EMRI signals with scalar-charge modifications from those affected by specific environmental effects. In the top-left panel, $\mathcal{M}(h^{\rm scalar}, h^{\rm DM})$ denotes the mismatch between EMRI waveforms with scalar radiation and those influenced by DM friction.
The top-right panel illustrates the difference between waveforms corrected by a scalar charge with $q_s = 0.01$ and those modified by accretion-disk effects, for a viscosity parameter $\alpha_{\rm disk} \in [0.01, 0.1]$ and a fixed scale-height parameter $h_0 = 0.05$. The bottom-left panel explores the dependence of the mismatch on the disk scale height $h_0 \in [0.01, 0.1]$, with fixed $\alpha_{\rm disk} = 0.04$ and scalar charge $q_s = 0.01$. The bottom-right panel considers the combined environmental scenario with both DM and accretion-disk effects, fixing $\alpha_{\rm disk} = 0.04$ and $\alpha_{\rm DM} = 1.65$, and varying the parameters $(h_0, e)$.

Among the environmental parameters, we find that the disk scale height $h_0$ has a more pronounced impact on the waveform mismatch than $\alpha_{\rm disk}$, which characterizes the disk viscosity. Hence, the bottom-right panel focuses on the mismatch as a function of the parameters $(h_0, e)$. From the top-left panel, it is evident that the deviation induced by DM friction becomes more distinguishable from scalar emission at lower values of the power-law index $\alpha_{\rm DM}$. In contrast, for a fixed $h_0 = 0.05$ in the bottom-left panel, scalar emission can be more easily distinguished from disk-induced effects by LISA. However, when both DM and accretion-disk effects are present, as shown in the bottom-right panel, the scalar dipole radiation becomes increasingly difficult to detect, regardless of the orbital eccentricity. This may be attributed to the fact that the dissipative forces from DM and disk friction partially cancel each other's influence, thereby reducing the overall waveform mismatch.
These findings underscore the importance of modeling scalar radiation in complex, non-vacuum environments. Accurate EMRI waveform modeling beyond the vacuum GR framework must account for such environmental degeneracies in order to ensure reliable parameter estimation and robust constraints on fundamental fields in future work.

\begin{table*}[htbp!]
\centering
\begin{tabular}{c|cccccccccc}
\hline
\hline
$e$ & $\sigma_M/M$ & $\sigma_\mu/\mu$   &$\sigma_{p_0}/p_0$
&$\sigma_{e_0}$
& $\sigma_{\alpha_{\text{DM}}}/\alpha_{\text{DM}}$
& $\sigma_{h_0}/h_0$  & $\sigma_{\Sigma_0}/\Sigma_0$
& $\sigma_{q_s}/q_s$  & $\sigma_{\Phi{_r,0}}/\Phi_{r,0}$
& $\sigma_{\Phi{_\phi,0}}/\Phi_{\phi,0}$
\\
\hline
0.05   &$1.59\text{e-4}$  &$7.13\text{e-2}$  &$8.75\text{e-4}$ &$4.57\text{e-4}$ & $1.44\text{e-3}$
& $0.16$  &$0.33$  & $6.36\text{e-1}$
&$7.57\text{e-1}$ &$4.57\text{e-1}$
\\
0.1   &$1.21\text{e-4}$  &$1.43\text{e-2}$  &$4.43\text{e-4}$ &$3.25\text{e-4}$ & $1.12\text{e-3}$
&$0.28$  &$0.76$  &$5.42\text{e-1}$
&$6.43\text{e-1}$ &$3.26\text{e-1}$
\\
0.2  &$8.85\text{e-5}$  &$7.62\text{e-3}$  &$1.27\text{e-4}$ &$2.46\text{e-4}$ & $8.12\text{e-4}$
&$0.53$  &$0.87$  &$2.13\text{e-1}$
&$3.08\text{e-1}$ &$1.12\text{e-1}$\\
\hline
$e$ & $\sigma_M/M$ & $\sigma_\mu/\mu$   &$\sigma_{p_0}/p_0$
&$\sigma_{e_0}$
& $-$
& $\sigma_{h_0}/h_0$ & $\sigma_{\Sigma_0}/\Sigma_0$
& $\sigma_{q_s}/q_s$  & $\sigma_{\Phi{_r,0}}/\Phi_{r,0}$
& $\sigma_{\Phi{_\phi,0}}/\Phi_{\phi,0}$
\\ \hline
0.05  &$5.63\text{e-5}$  &$6.03\text{e-3}$  &$1.92\text{e-4}$ &$3.76\text{e-4}$ & $-$
&$0.13$  &$0.11$  &$2.69\text{e-1}$
&$4.75\text{e-1}$ &$4.10\text{e-1}$\\
0.1  &$4.25\text{e-5}$  &$4.35\text{e-3}$  &$1.18\text{e-4}$ &$2.46\text{e-4}$ & $-$
&$0.25$  &$0.37$  &$1.34\text{e-1}$
&$3.16\text{e-1}$ &$3.24\text{e-1}$ \\
0.2  &$2.67\text{e-5}$  &$2.28\text{e-3}$  &$1.08\text{e-4}$ &$1.78\text{e-4}$ & $-$
&$0.46$  &$0.73$  &$1.27\text{e-1}$
&$2.35\text{e-1}$ &$2.53\text{e-1}$ \\
\hline
$e$ & $\sigma_M/M$ & $\sigma_\mu/\mu$   &$\sigma_{p_0}/p_0$
&$\sigma_{e_0}$
&$\sigma_{\alpha_{\text{DM}}}/\alpha_{\text{DM}}$
& $-$ & $-$
& $\sigma_{q_s}/q_s$  & $\sigma_{\Phi{_r,0}}/\Phi_{r,0}$
& $\sigma_{\Phi{_\phi,0}}/\Phi_{\phi,0}$  \\ \hline
0.05  &$1.96\text{e-5}$  &$5.33\text{e-3}$  &$3.46\text{e-4}$ &$3.29\text{e-4}$  &$1.43\text{e-3}$
&$-$  &$-$  &$2.76\text{e-2}$
&$1.31\text{e-1}$ &$2.28\text{e-2}$ \\
0.1  &$1.12\text{e-5}$  &$4.28\text{e-3}$  &$2.34\text{e-4}$ &$2.57\text{e-4}$  &$1.25\text{e-3}$
&$-$  &$-$  &$1.84\text{e-2}$
&$1.25\text{e-1}$ &$2.35\text{e-2}$
\\
0.2  &$1.05\text{e-5}$  &$2.46\text{e-3}$  &$1.48\text{e-4}$ &$1.38\text{e-4}$  &$1.18\text{e-3}$
&$-$  &$-$  &$1.07\text{e-2}$
&$5.78\text{e-2}$ &$1.68\text{e-2}$ \\
\hline
\hline
\end{tabular}
\caption{Measurement errors from FIM for the binary parameters are listed.
The binary system has component masses $(M=10^{6}M_\odot,\mu= 30M_\odot)$,
scalar charge $(q_s=0.1)$, the accretion-disk central surface density $(\Sigma_0=5.25\times10^4 \rm{g/cm^3})$ and aspect ratio $(h_0=0.025)$, the power law index of DM $(\alpha_{\text{DM}}=1.7)$, initial orbital phases $(\Phi_{r,0}=\Phi_{\phi,0}=1.0)$ and orbital semi-latus rectum $(p_0=10.0)$. The inspiral dration is fixed to four years and the luminosity distance is adjusted to fix SNR of EMRIs signal as $30$. The blanks in the table
denote to the absent of environmental effect, such as the transverse lines below ``$\sigma_{\alpha_{\text{DM}}}/\alpha_{\text{DM}}$'' means that  EMRIs without dynamic friction of DM halo.
}\label{tab:fim:error}
\end{table*}

\begin{figure*}[htb!]
\centering
\includegraphics[width=0.87\paperwidth]{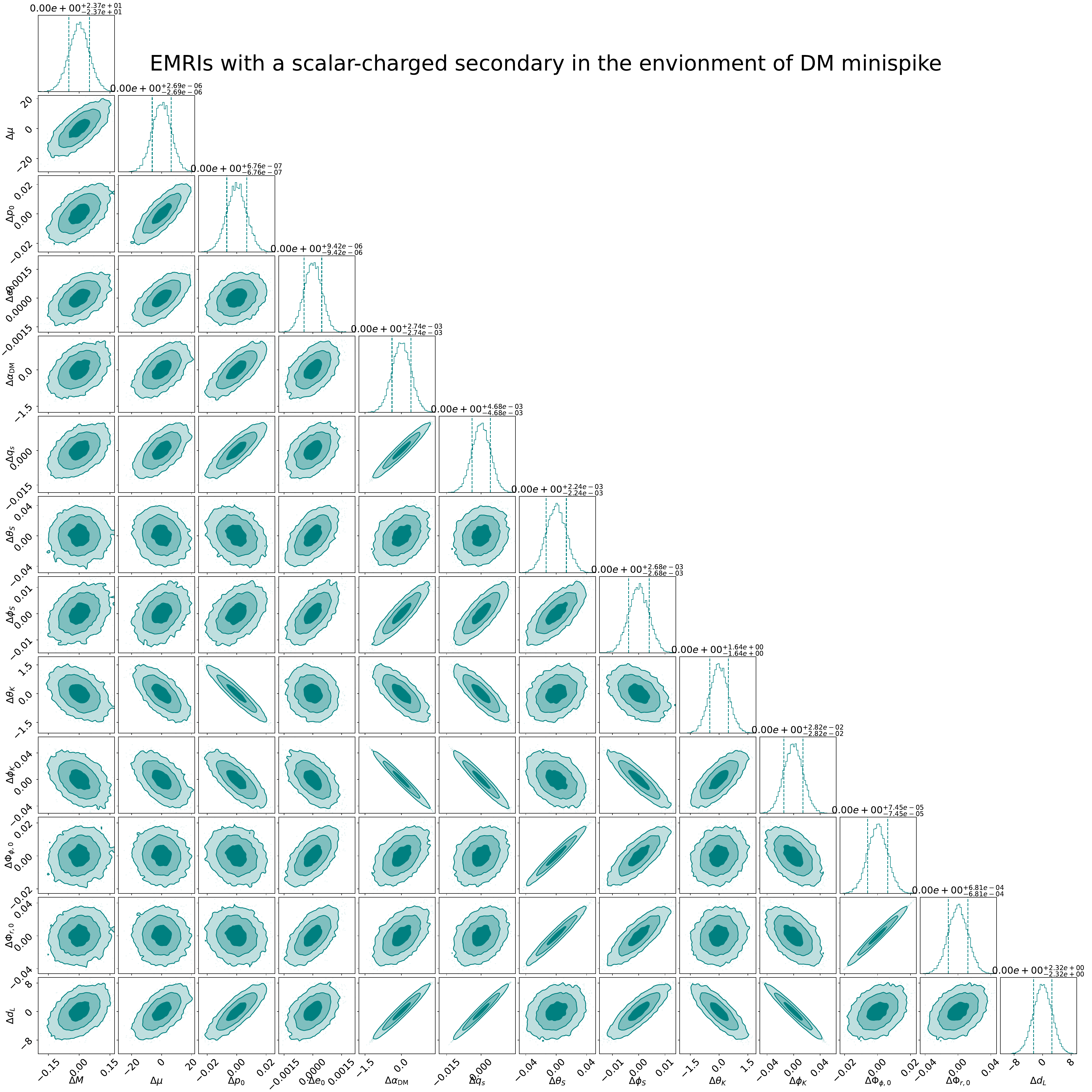}
\caption{Corner plot of the posterior probability distributions for the mass of the secondary, the mass of the primary, the initial orbital eccentricity, the initial semi-latus rectum, the power-law index of the DM minispike, the scalar charge, the directional parameters of the source and orbital angular momentum, the accumulated orbital phases, and the luminosity distance. The fiducial parameters are $(M=10^6M_\odot,\ \mu=30M_\odot,\ p_0=13,\ e_0=0.1,\ \alpha_{\text{DM}}=1.7,\ q_s=0.1,\ \theta_S=\pi/3,\ \phi_S=\pi/4,\ \theta_K=\pi/4,\ \phi_K=\pi/5,\ \Phi_{\phi,0}=1.0,\ \Phi_{r,0}=1.0,\ d_L=1~\textrm{Gpc})$, and the constraints are inferred from a four-year observation. Vertical dashed lines indicate the $1\sigma$ credible intervals for each parameter, while the contours correspond to the $68\%$, $95\%$ and $99\%$ credible regions.
} \label{fig:cornerplot:env:DM}
\end{figure*}

\begin{figure*}[htb!]
\centering
\includegraphics[width=0.87\paperwidth]{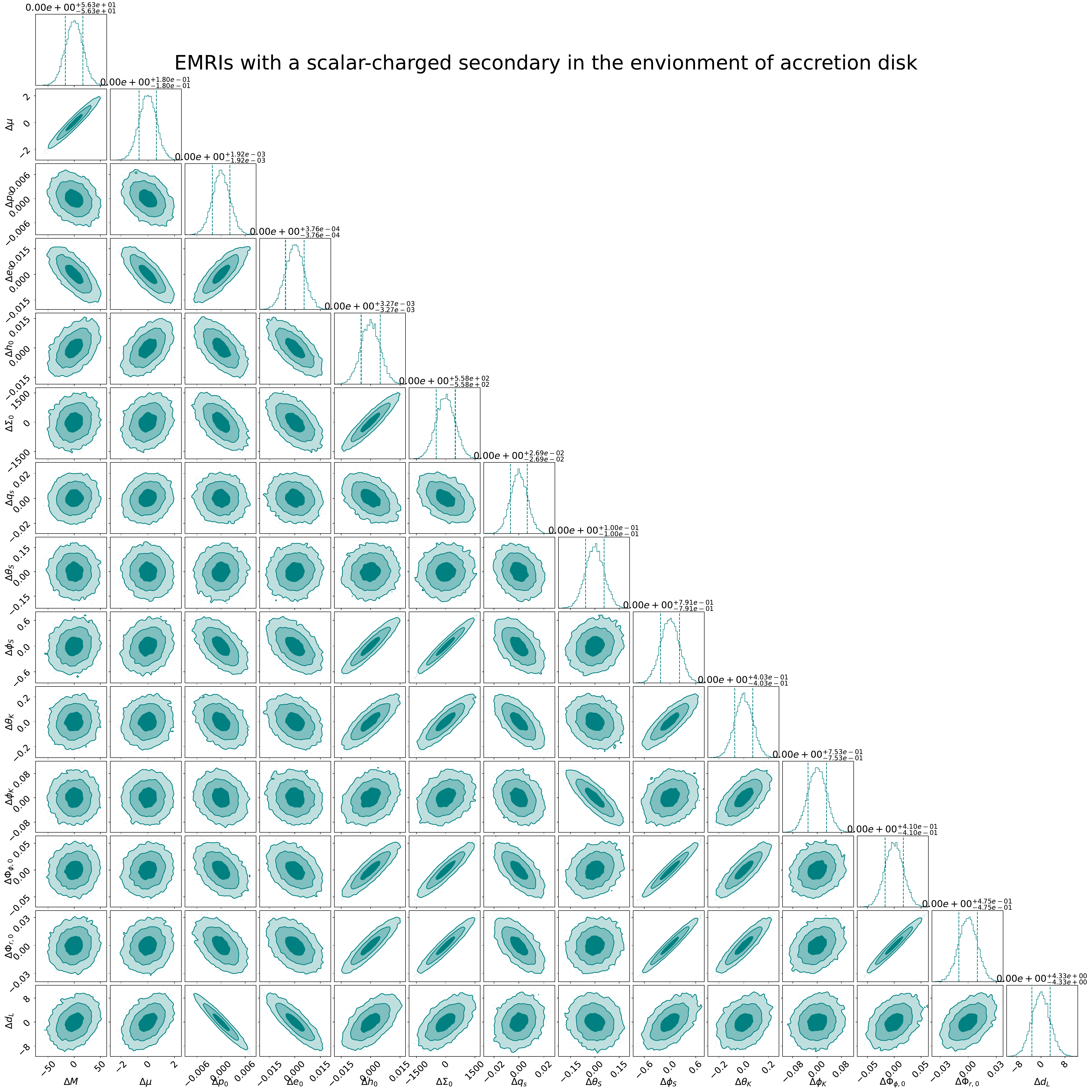}
\caption{Corner plot of the posterior probability distributions for the component masses $(M,\mu)$, the initial orbital parameters $(p_0,e_0)$, the accretion-disk parameters $(h_0,\Sigma_0)$, the scalar charge $q_s$, the sky-location angles $(\theta_S,\phi_S)$, the orientation angles $(\theta_K,\phi_K)$ of orbital angular momentum, the initial orbital phases $(\Phi_{\phi,0},\Phi_{r,0})$, and the luminosity distance $d_L$. The posteriors are computed for a fiducial EMRI with $(M=10^6 M_\odot,\ \mu=30 M_\odot,\ p_0=13,\ e_0=0.1,\ h_0=0.025,\ \Sigma_0=5.25\times10^3~{\rm g/cm^{-3}},\ q_s=0.1,\ \theta_S=\pi/3,\ \phi_S=\pi/4,\ \theta_K=\pi/4,\ \phi_K=\pi/5,\ \Phi_{\phi,0}=1.0,\ \Phi_{r,0}=1.0,\ d_L=1~\textrm{Gpc})$, assuming a four-year observation. Vertical dashed lines denote the $1\sigma$ credible intervals for each parameter, and the contours correspond to the $68\%$, $95\%$, and $99\%$ credible regions.} \label{fig:cornerplot:env:disk}
\end{figure*}

\begin{figure*}[htb!]
\centering
\includegraphics[width=0.87\paperwidth]{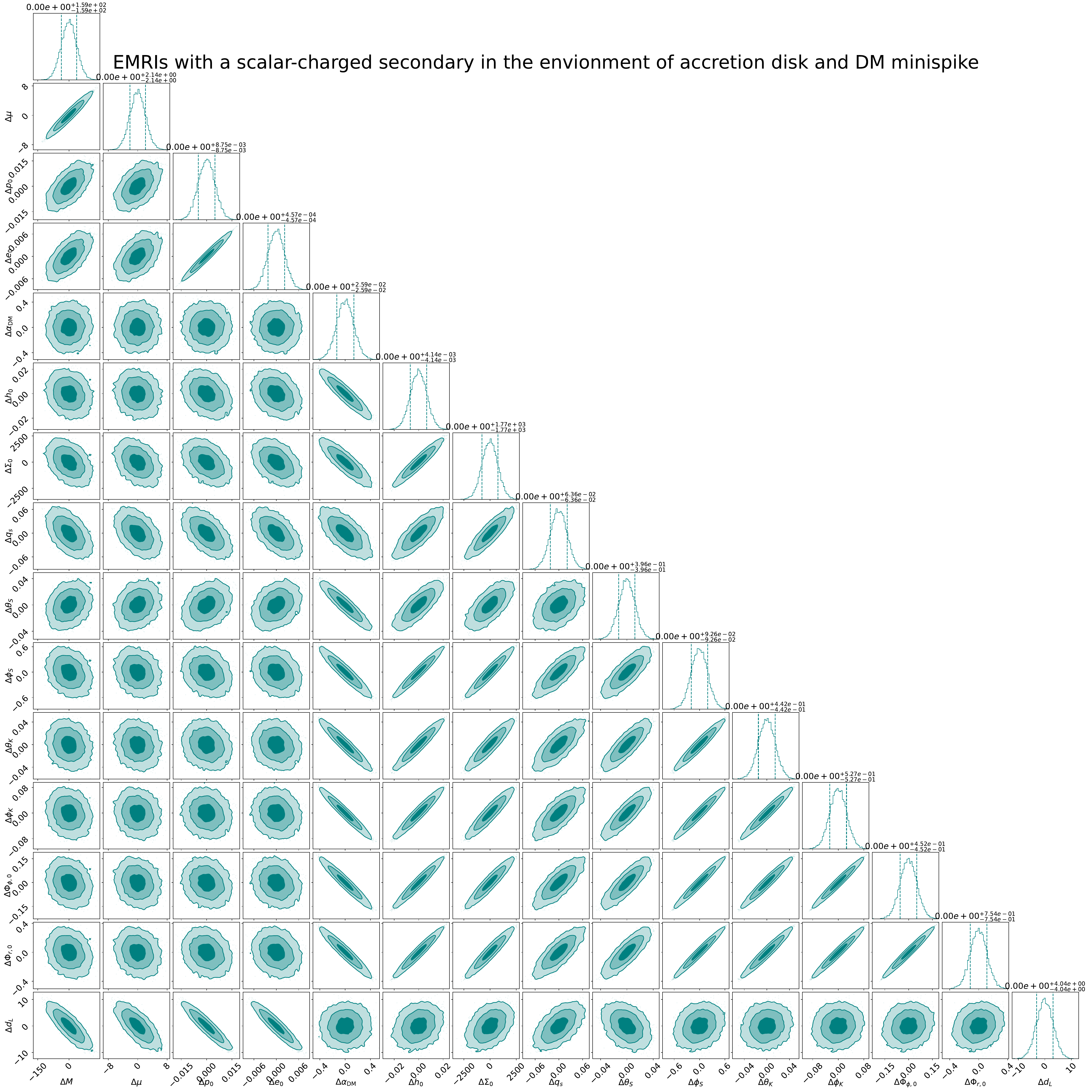}
\caption{Corner plot displaying the posterior probability distributions for the component masses $(M,\mu)$, the initial orbital parameters $(p_0,e_0)$, the DM-minispike power-law index $\alpha_{\rm DM}$, the accretion-disk parameters $(h_0,\Sigma_0)$, the scalar charge $q_s$, the sky-location angles $(\theta_S,\phi_S)$, the orientation angles $(\theta_K,\phi_K)$ of orbital angular momentum, the initial orbital phases $(\Phi_{\phi,0},\Phi_{r,0})$, and the luminosity distance $d_L$. The posteriors are obtained for a fiducial EMRI with $(M=10^6 M_\odot,\ \mu=30 M_\odot,\ p_0=13,\ e_0=0.1,\ \alpha_{\rm DM}=1.7,\ h_0=0.025,\ \Sigma_0=5.25\times10^3~{\rm g\,cm^{-3}},\ q_s=0.1,\ \theta_S=\pi/3,\ \phi_S=\pi/4,\ \theta_K=\pi/4,\ \phi_K=\pi/5,\ \Phi_{\phi,0}=1.0,\ \Phi_{r,0}=1.0,\ d_L=1~\textrm{Gpc})$, based on a four-year observation. Vertical dashed lines mark the $1\sigma$ credible intervals for each parameter, and the contours indicate the $68\%$, $95\%$, and $99\%$ credible regions.} \label{fig:cornerplot:env:tot}
\end{figure*}

At the end of this section, we summarize in Table~\ref{tab:fim:error} the measurement precision of the source parameters inferred from a LISA observation. For all cases, we assume a four-year EMRI signal including environmental corrections, and we freely adjust the luminosity distance so that the SNR is fixed at $\rho=30$. Three environmental configurations are considered: (i) a DM minispike interacting with the secondary, (ii) an accretion disk surrounding the MBH, and (iii) an EMRI embedded in the environment of both a DM minispike and an accretion disk.

Overall, the scalar charge can be constrained with an absolute uncertainty of order $\sim 10^{-2}$, with the relative error reaching $\sim 10^{-1}$ in the most complex environments. A higher initial orbital eccentricity systematically improves the bounds on the scalar charge, irrespective of the environmental configuration. The measurement errors of the remaining parameters are moderately degraded relative to the vacuum case~\cite{Babak:2017tow,Fan:2020zhy}, which may be attributed to the fact that the current waveform model does not include relativistic couplings among the different environmental effects. This degradation may be alleviated once fully relativistic environmental corrections are incorporated in future work~\cite{Cardoso:2022whc,Duque:2023seg,Spieksma:2024voy,Cardoso:2021wlq}.

In Fig.~\ref{fig:cornerplot:env:DM}, we also perform the analysis of the correlation among different source parameters using EMRIs signal formed in the DM halo, in which the corner plot is inferred from off-diagonal elements of covariance matrix.
One can find that the bound level of scalar charge not only depends on the measurement of intrinsic parameters, but also relating to the sky location of the source. There is a strong positive correlation between scalar charge $q_s$ and other parameters, except the angle $\theta_K$.
The following case accounts for the more true astrophysical effect (accretion disk) existing around a MBH, which generates a non-neglected imprint on EMRIs signal. Fig.~\ref{fig:cornerplot:env:disk} illustrates the case for the environments of an acctetion disk, characterized by parameters  $(h_0,\Sigma_0)$. In this configuration, the correlation of scalar charge and other parameters is significantly decreased.
However, when both the effects of DM and accretion disk in EMRIs are included in Fig.~\ref{fig:cornerplot:env:tot}, the measurement of scalar charge with LISA are positive related with the environmental parameters and other intrinsic quantities.
Therefore, these findings underscore the importance of accurately modeling the environmental effects in EMRIs waveform from beyond-GR theories to enable robust test of fundamental physics with LISA.
A comprehensive waveform model that accounts for relativistic interactions between accretion disk and DM is essential to disentangling the potential signatures of non-GR.

\section{Discussion}\label{discussion}
In this paper, we computed the modification of two environmental effect on EMRIs waveform to assess how the environments influence on the constraint of scalar charge, including the accretion disk and DM dynamic friction. In particular, we employ a hybrid scheme to compute the quadrupole waveform by the inspiraling trajectories corrected
the accretion-disk and fluxes modified by the DM and scalar charge.
Then we assess the difference of various environmental effects on EMRIs waveforms with correction and without of the scalar emission,
present a constraint on scalar charge using FIM and the correlation among different source parameters.
According to the mismatch analysis, under the appropriate parametric setting, EMRIs signal with correction of scalar charge in beyond-vacuum spacetime can be distinguished from that in the pure vacuum spacetime. When there is only the DM minispike in EMRIs, the larger power-low index $\alpha_{\text{DM}}$ can help us to discern the effect of scalar charge. If the secondary is locating in the accretion disk and influenced by dynamic frictional force,  kinematic viscosity $\alpha_{\rm disk}$ and aspect ratio $h_0$ do not generate a significant effect on the discrimination of scalar charge.
With the EMRIs signal modified by different environments, LISA can detect the scalar charge within a relative error of $\sim0.1$, the bound also depends on the measurement errors of other intrinsic parameters. If considering EMRIs in the environments consisting of DM and accretion disk, the correlation of scalar charge and other source parameters are mainly positive. This can show that accurately modeling these environmental effects on EMRIs fluxes and waveforms is necessary to test fundamental theories.

Our work presents the first assessment of scalar charge measurability with LISA, incorporating two environmental effects in Schwarzschild spacetime.
The current waveform template relies on a simplified assumption: environmental effects solely alter EMRI energy fluxes or trajectories, neglecting modifications to orbital fundamental frequencies by astrophysical processes. Such modifications require rigorous treatment in future analyses~\cite{Li:2025ffh,Gliorio:2025cbh,Rahman:2025mip}.
Furthermore, astrophysical MBHs inherently possess spin. Subsequent work must address how environmental effects influence EMRI fluxes in a rotating spacetime.
Our FIM approach for estimating environmental effects with space-based detectors remains simplistic. Bayesian inference methods, particularly Markov Chain Monte Carlo (MCMC), offer a more robust alternative for GW data analysis, mitigating the FIM's limitations.

\section*{Acknowledgements}
This work is funded by the National Natural Science Foundation of China with Grants No. 12347140 and No. 12405059. C.Ye is funded by
the Sichuan University of Science $\&$ Engineering Program (No. 2024RC031).

%\bibliographystyle{apsrev}
%\bibliography{JN1}

\end{document}